\documentclass[seceq,supplement]{ptptex}
\usepackage{graphicx,wrapfig}
\newlength{\minitwocolumn}
\newlength{\minispace}
\newlength{\figsmall}
\newlength{\figwidth}
\newlength{\figmid}
\newlength{\figlarge}
\title{
Finite-Temperature Mott Transition in Two-Dimensional Frustrated Hubbard Models
}

\author{
Takuma \textsc{Ohashi} $^{1,2)}$,
Tsutomu \textsc{Momoi} $^{2)}$, \\
Hirokazu \textsc{Tsunetsugu} $^{3)}$ 
and 
Norio \textsc{Kawakami} $^{4)}$
}

\inst{
$^1$Department of Physics, Osaka University, 
Toyonaka, Osaka 560-0043, Japan. \\
$^2$Condensed Matter Theory Laboratory, RIKEN,
Wako, Saitama 351-0198, Japan. \\
$^3$Institute for Solid State Physics,
University of Tokyo, 
Kashiwa, Chiba 277-8581, Japan. \\
$^4$Department of Physics, Kyoto University,
Kyoto 606-8502, Japan
}

\abst{
We investigate the Hubbard model on two typical frustrated lattices
in two dimensions, 
the kagom{\'e} lattice and the anisotropic triangular lattice, 
by means of the cellular dynamical mean field theory. 
We show that the metallic phase is stabilized up to fairly large Hubbard interactions under strong geometrical frustration in both cases, which
results in heavy fermion behavior and several anomalous properties 
 around the Mott transition point. In particular,
for the anisotropic triangular lattice, we find novel reentrant behavior 
in the Mott transition in the moderately frustrated parameter regime, 
which is caused by the competition between Fermi-liquid formation and 
magnetic correlations. It is  demonstrated that the reentrant behavior 
is a generic feature inherent in 
the Mott transition with intermediate geometrical frustration, 
and indeed in accordance with recent experimental 
findings for organic materials. 
}

\begin{document}
\ifx\href\undefined\else\hypersetup{linktocpage=true}\fi 
\maketitle

\section{Introduction\label{sec:intro}}
Geometrical frustration has attracted much interest in strongly correlated 
electron systems. Among a number of intriguing phenomena, 
the observation of heavy fermion behavior in $\mathrm{LiV_2O_4}$ 
\cite{kondo97,jonsson07} with the pyrochlore lattice structure  
has activated theoretical studies of electron correlations 
with geometrical frustration. Also, the discovery of superconductivity 
in the triangular-lattice oxide
$\mathrm{Na}_x\mathrm{CoO}_2 \cdot y\mathrm{H}_2\mathrm{O}$ \cite{takada03} 
and the $\beta$-pyrochlore osmate $A \mathrm{Os}_2\mathrm{O}_6$
($A=$K \cite{yonezawa04k}, Rb, \cite{yonezawa04rb} Cs \cite{yonezawa04cs}) 
has stimulated further investigations of frustrated 
electron systems. These intensive studies have 
revealed new aspects of the Mott transition for 
geometrically frustrated electrons. In particular, an experimental
demonstration of the spin liquid ground state in the Mott insulating phase 
in the organic material 
$\kappa$-(BEDT-TTF)$_2\mathrm{Cu}_2\mathrm{CN}_3$ \cite{shimizu03} 
with triangular lattice structure,
raises a theoretical challenge 
in the physics of geometrically frustrated electron systems. 
Other materials found recently, such as 
the pyrochlore Kondo lattice compound 
$\mathrm{Pr}_2\mathrm{Ir}_2\mathrm{O}_7$ \cite{nakatsuji06} 
and the hyperkagom{\'e} compound 
$\mathrm{Na}_4\mathrm{Ir}_3\mathrm{O}_8$ \cite{okamoto07}, 
also provide new examples of frustrated electron 
systems with nonmagnetic ground state.
In this paper, we will investigate the two fundamental
frustrated systems, the kagom{\'e} lattice 
and the anisotropic triangular lattice, which we will
briefly explain below.

\subsection{Kagom{\'e} lattice}

The kagom{\'e} lattice is one of typical frustrated systems 
and it shares some essential properties of the pyrochlore lattice. 
Antiferromagnetic spin systems on this lattice 
have been intensively studied and many unusual properties 
have been found \cite{misguich04,shores05}. 
Theoretical studies of the $S=1/2$ Heisenberg antiferromagnet 
has suggested the realization of a nonmagnetic ground state 
and the existence of anomalous singlet excitations 
within the singlet-triplet gap due to strong frustration. 
Therefore, it is natural to ask what kind 
of quasiparticle dynamics 
these unusual properties induce if electrons become itinerant. 
An interesting example of the itinerant kagom{\'e} systems 
may be a superconducting compound
$\mathrm{Na}_x\mathrm{CoO}_2 \cdot y\mathrm{H}_2\mathrm{O}$. 
It has been suggested that an effective model of this material 
can be regarded as a correlated electron system on the kagom{\'e} lattice 
by properly considering anisotropic hopping
matrix elements of the cobalt $3d$ orbitals \cite{koshibae03}. 
Also, the hyperkagom{\'e} compound 
$\mathrm{Na}_4\mathrm{Ir}_3\mathrm{O}_8$ is 
a three dimensional analog of the kagom{\'e} lattice electron system. 
The issue of electron correlations for the kagom{\'e} lattice 
was addressed recently in the studies  
by using the fluctuation-exchange 
(FLEX) approximation \cite{imai03} and quantum Monte Carlo (QMC) method \cite{bulut05}.  These studies focused on electron 
correlations in the metallic regime, and the nature of the
Mott transition has not been clarified. 
We shall investigate in this paper the kagom{\'e} lattice electron system
with particular emphasis on the Mott transition under the
influence of strong frustration. 

\subsection{Triangular lattice}

Another remarkable example is a triangular electron system with 
strong correlations, in which the metal-insulator transition
has intensively studied. This has particularly been stimulated 
by various interesting phenomena recently found 
in organic materials $\kappa$-(BEDT-TTF)$_2X$
around the Mott transition, 
such as a spin liquid state, 
unconventional superconductivity, etc
\cite{lefebvre00,shimizu03,kagawa04}. 
A possible nonmagnetic ground state is found in the triangular lattice Hubbard model 
by the path integral renormalization group study, 
which provides a powerful numerical treatment of 
the frustrated electron systems \cite{kashima01}, 
and the correlated electrons on the anisotropic triangular lattice have
been intensively studied so far 
\cite{imai02,onoda03prb,onoda03jpsj,parcollet04,yokoyama06,mizusaki06,
kyung06prl,kyung07,kancharla07,koretsune07}. 
The effects of geometrical frustration on finite-temperature ($T$) 
Mott transition, however, have not yet been sufficiently understood. 
One of the interesting and nontrivial features of the finite-$T$ 
Mott transition is a reentrant behavior observed in the frustrated 
organic material $\kappa$-(BEDT-TTF)$\mathrm{_2Cu[N(CN)_2]Cl}$
under pressure \cite{lefebvre00,kagawa04}. 
With lowering temperature, 
it once undergoes a transition from Mott 
insulator to metal, and then reenters the paramagnetic insulating
phase at a much lower temperature.
This reentrant behavior is 
quite different from the nonreentrant behavior of Mott transition 
in the three dimensional systems, 
such as $\mathrm{V_2O_3}$, and is expected to be a new
aspect of the geometrical frustration and possibly low-dimensionality.
We will address this problem in this paper.

\subsection{Theoretical approach}

In order to investigate the Mott transition 
in the kagom{\'e} and triangular electron systems, 
we need advanced theoretical methods. Among many 
approaches for correlated electron systems, the dynamical mean 
field theory 
(DMFT) \cite{metzner89,georges96,kotliar06} 
has given 
substantial theoretical progress in understanding the Mott transition \cite{imada98}, 
and it has also clarified various interesting phenomena 
\cite{georges96,bulla99}
in the strongly correlated electron systems, 
such as magnetism 
\cite{jarrell92,rozenberg95,momoi98,si01,sun03,zitzler04,sakai07}, 
heavy fermion formation 
\cite{jarrell93,mutou94,saso96,ohashi04,ohashi05,arita07}, 
orbital physics in the multiband systems 
\cite{han98,florens02,koga02,koga04,sato04,medici05,arita05,koga05,inaba07}, etc. 
Recently, DMFT has been also applied to cold atoms in an optical lattice 
\cite{helmes08,titvinidze08,higashiyama08,koga08}, 
some inhomogeneous systems \cite{potthoff99,okamoto04,freericks06}, and 
photo-excited semiconductors \cite{tomio05,ogawa07}. 
However, DMFT does not take account of spatially extended correlations, which should 
be included for the systems under consideration in this paper.
Therefore, it is desirable to study the Mott transition 
by employing another appropriate method 
which properly incorporate spatially extended correlations and 
geometrical frustration. 
Cluster extensions of DMFT \cite{hettler98,lichtenstein00,kotliar01,okamoto03,maier05} 
or the self energy functional approach \cite{potthoff03}
are candidates for this purpose. 
Recently developed diagrammatic extensions of DMFT \cite{toschi07,rubtsov08}
might also treat geometrical frustration by incorporating 
the $\mathbf{k}$-dependence of the self-energy. 
Here we will use a cluster extension of DMFT, 
the cellular dynamical mean field theory 
(CDMFT) \cite{lichtenstein00,kotliar01}.

\subsection{Purpose of the paper}

In this paper, we give a brief review of our recent studies on
the Mott transitions in the Hubbard model on the geometrically 
frustrated kagom{\'e} \cite{ohashi06,ohashi07jpcm,ohashi07jmmm} 
and anisotropic triangular lattices \cite{ohashi08}
by means of CDMFT combined 
with QMC \cite{hirsch86}. We investigate these models 
separately to discuss properties characteristic of each system, and then
deduce common properties inherent in frustrated electron systems. 
In both models, we find that the metallic phase is stable up to 
fairly large  Hubbard interactions under strong geometrical frustration, 
giving rise to the heavy fermion behavior near the Mott transition.
In the kagom{\'e} lattice, several anomalous properties 
of spin correlation functions, such as nonmonotonic 
temperature dependence, emerge around the Mott transition. 
For the anisotropic triangular lattice, we discover more
striking behavior in the Mott transition. Namely, in moderately 
frustrated cases, 
 the finite-$T$ Mott transition shows a reentrant behavior, which is 
consistent with experiments in some organic materials \cite{lefebvre00,kagawa04}. 
We demonstrate that the reentrant behavior is a characteristic feature 
inherent in the Mott transition with geometrical frustration, 
and thus can be experimentally observed in various frustrated electron systems. 

The paper is organized as follows. 
In the next section, we introduce the model Hamiltonian and 
briefly explain the framework of CDMFT. 
We first study the Mott transition  in 
the kagom{\'e} lattice Hubbard model and elucidate
some anomalous properties appearing near the transition 
point in \S \ref{sec:kagome}. 
In \S \ref{sec:tri}, we then investigate the reentrant Mott transition 
on the anisotropic triangular lattice. 
A brief summary is given in the last section. 

\section{Model and method\label{sec:method}}
We consider the standard Hubbard model 
on the kagom{\'e} lattice (see Fig. \ref{fig:kagome}(a)) and the
anisotropic triangular lattice (see Fig. \ref{fig:lattice}), 
\begin{align}
H=   \sum_{i,j,\sigma} t_{ij}
     c_{i\sigma }^\dag c_{j\sigma}
   + U \sum_{i} n_{i\uparrow} n_{i\downarrow} ,
\label{eqn:hm}
\end{align}
with $n_{i\sigma}=c_{i\sigma}^\dag c_{i\sigma}$,
where $c_{i\sigma }^\dag$ ($c_{i\sigma}$) creates (annihilates)
an electron with spin $\sigma$ at site $i$. 
Here, $i, j=1,2,\cdots,N$, and $N$ is the total number of sites. 
The hopping matrix element and the Hubbard interaction are denoted as
$t_{ij}$ and $U$, respectively. The explicit form of $t_{ij}$ will be
given for each model below. Note that in both models, a triangular 
structure of the unit cell, which is a source of strong frustration, 
 plays a crucial role in controlling the nature of 
Mott transition. Therefore, theoretical methods 
beyond DMFT are necessary to incorporate spatially extended electron 
correlations. To this end,  we here use CDMFT, 
which has been successfully applied to frustrated systems such as
the Hubbard model on the triangular lattice \cite{parcollet04,kyung06prl,kyung07} and the
kagom{\'e} lattice \cite{ohashi06}.

\subsection{Cellular dynamical mean field theory}

\begin{figure}[tb]
\begin{center}
\includegraphics[clip,,width=\figlarge]{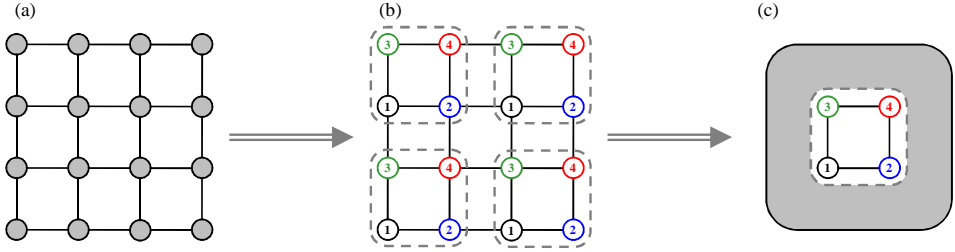}
\end{center}
\caption{\label{fig:cdmft}(Color online) 
Sketch of the original lattice (a), 
superlattice (b) and effective cluster model (c) 
for the cluster size $N_c=4$. 
}
\end{figure}

In CDMFT, the original lattice is regarded as 
a superlattice consisting of small clusters, 
as shown in Fig. \ref{fig:cdmft} (b). 
Here, we define the  cluster size as $N_c$. 
The model (\ref{eqn:hm}) is rewritten as, 
\begin{align}
H=   \sum_{l,m,\gamma,\delta,\sigma} t_{\gamma,\delta}\left( l, m \right)
     c_{l \gamma \sigma }^\dag c_{m \delta \sigma}
   + U \sum_{l,\gamma} n_{l \gamma \uparrow} n_{l \gamma \downarrow} ,
\label{eqn:hm_sl}
\end{align}
where $l$ and $m$ are cluster indices, 
$l, m = 1, 2, \cdots , N/N_c$, and
$\gamma$ and $\delta$ are sublattice indices, 
$\gamma, \delta = 1, 2, \cdots , N_c$. 
Using a standard DMFT procedure with 
the inter-cluster hopping scaled as 
$t_{\gamma,\delta}(l,m) \to t_{\gamma,\delta}(l,m)/\sqrt{d}$, 
the original model is then mapped onto an effective cluster model 
consisting of a cluster coupled to 
the self-consistently determined medium, 
as illustrated in Fig. \ref{fig:cdmft} (c). 
The corresponding action reads
\begin{align}
S_\mathrm{eff} & = 
\int_0^\beta d\tau \int_0^\beta d\tau' \sum_{\gamma,\delta,\sigma}
c_{\gamma\sigma}^\dag \left( \tau \right)
\left( \mathcal{G}^{-1} \right)
_{\gamma \delta \sigma}
\left( \tau - \tau' \right)
c_{\delta\sigma} \left( \tau' \right)
\nonumber \\
& + U \int_0^\beta d\tau 
\sum_{\gamma}
n_{\gamma \uparrow} \left( \tau \right) 
n_{\gamma \downarrow} \left( \tau \right). 
\label{eqn:action}
\end{align}
Here, $\beta=1/T$. 
Given the Green's function for the effective medium, 
$\hat{\mathcal{G}}_{\sigma}$, 
we can compute the cluster Green's function $\hat{G}_{\sigma}$
by solving the effective cluster model with QMC method \cite{hirsch86}, 
and then we obtain the cluster self-energy $\hat{\Sigma}_{\sigma}$. 
Here, $\hat{\mathcal{G}}_{\sigma}$, $\hat{G}_{\sigma}$, 
and $\hat{\Sigma}_{\sigma}$ are $N_c \times N_c$ matrices. 
In order to reduce errors due to finite time slices in QMC, 
we exploit an interpolation scheme based on a high-frequency expansion
of the discrete imaginary-time Green's function \cite{oudovenko02}. 
The effective medium $\hat{\mathcal{G}}_{\sigma}$ 
is then updated by the Dyson equation, 
\begin{align}
\hat{\mathcal{G}}_{\sigma}^{-1} \left( i\omega_n \right) &= 
\left[ \frac{N_c}{N} \sum_\mathbf{\tilde{k}} 
\hat{g}_\sigma \left( \mathbf{\tilde{k}}: i\omega_n \right) \right]^{-1}
+ \hat{\Sigma}_{\sigma} \left( i\omega_n \right) , 
\label{eqn:bath} \\
\hat{g}_\sigma \left( \mathbf{\tilde{k}} : i\omega_n \right) &=
\left[ i\omega_n + \mu - \hat{t} \left( \mathbf{\tilde{k}} \right)
- \hat{\Sigma}_\sigma \left( i\omega_n \right) \right ]^{-1}, 
\label{eqn:gf}
\end{align}
where $\mu$ is the chemical potential and 
$\hat{t} ( \mathbf{\tilde{k}} )$ is the 
Fourier-transformed hopping matrix for the superlattice, 
\begin{align}
t_{\gamma \delta} \left( \mathbf{\tilde{k}} \right) = \frac{N_c}{N}
\sum_{l,m} e^{-\mathbf{\tilde{k}} \cdot (\mathbf{r}_l - \mathbf{r}_m)}
t_{\gamma \delta} \left( l,m \right). 
\label{eqn:hopping}
\end{align}
Here the summation of $\mathbf{\tilde{k}}$ is taken over the reduced Brillouin zone 
of the superlattice. 

\subsection{Wave-vector dependent properties}
Within CDMFT, the single-electron Green's function 
for wave vector $\mathbf{k}$ is given as, 
\begin{align}
G_\mathbf{k}\left( i\omega_n \right) = \frac{1}{N_c}
\sum_{\gamma,\delta}
e^{i\mathbf{k} \cdot \left( \mathbf{r}_\gamma - \mathbf{r}_\delta \right)}
\left[ i\omega_n + \mu - \hat{t} \left( \mathbf{k} \right)
  - \hat{\Sigma} \left( i\omega_n \right)
\right ]^{-1}_{\gamma\delta},
\label{eqn:green}
\end{align}
where $\mathbf{k}$ is the wave vector in the original Brillouin zone
and $\mathbf{r}_\gamma$, $\mathbf{r}_\delta$ label cluster sites \cite{kyung06prb}. 
We calculate the imaginary time Green's 
function $G_\mathbf{k}\left( \tau \right)$ and obtain the spectrum
$A_\mathbf{k}\left( \omega \right) = -\mathrm{Im} G_\mathbf{k}\left(
\omega + i0 \right)/\pi$ using the maximum entropy method (MEM) \cite{jarrell96}.

It is also possible to compute the wave-vector dependence of 
various two-electron Green's functions 
with including vertex corrections. 
Here, we investigate the $\mathbf{q}$-dependent 
static spin susceptibility defined as, 
\begin{align}
\chi_{\gamma \delta}(\mathbf{\tilde{q}}) = 
\frac{N_c}{N} \int_0^{\beta} d \tau 
\sum_{\mathbf{\tilde{k}},\mathbf{\tilde{k}}'} 
\left \langle 
c_{\mathbf{\tilde{k}}\gamma\uparrow}^\dag               \left ( \tau \right ) 
c_{\mathbf{\tilde{k}}+\mathbf{\tilde{q}}\gamma\downarrow}       \left ( \tau \right )
c_{\mathbf{\tilde{k}}'+\mathbf{\tilde{q}}\delta\downarrow}^\dag \left ( 0 \right )
c_{\mathbf{\tilde{k}}'\delta\uparrow}                   \left ( 0 \right )
\right \rangle .
\label{eqn:chi}
\end{align}
In order to incorporate the vertex correction into 
the susceptibility, 
we consider the two-electron Green's function 
in the effective cluster model (\ref{eqn:action}), 
\begin{align}
C_{\gamma \delta} \left( i\omega_l,i\omega_m \right) &= \frac{1}{\beta}
\int_0^\beta \int_0^\beta \int_0^\beta \int_0^\beta
d\tau_1 d\tau_2 d\tau_3 d\tau_4 \nonumber \\
&\times
e^{-i\omega_l \left( \tau_1 - \tau_2 \right)}
e^{-i\omega_m \left( \tau_3 - \tau_4 \right)}
C_{\gamma \delta} \left( \tau_1,\tau_2,\tau_3,\tau_4 \right), \\
C_{\gamma \delta} \left( \tau_1,\tau_2,\tau_3,\tau_4 \right) &= 
\left \langle \mathrm{T}_\tau
c_{\gamma\uparrow}^\dag   \left ( \tau_1 \right ) 
c_{\gamma\downarrow}      \left ( \tau_2 \right )
c_{\delta\downarrow}^\dag \left ( \tau_3 \right )
c_{\delta\uparrow}        \left ( \tau_4 \right )
\right \rangle .
\label{eqn:cf}
\end{align}
We first calculate the cluster two-electron Green's function 
(\ref{eqn:cf}) by QMC and 
 extract the vertex function 
$\Gamma_{\gamma\delta}\left( i\omega_l,i\omega_m \right)$ 
via the Bethe-Salpeter equation, 
\begin{align}
\hat{\Gamma} = {\hat{C^0}}^{-1} - \hat{C}^{-1}, 
\label{eqn:vf}
\end{align}
where $C^0$ is the bare cluster two-electron Green's function, 
\begin{align}
C^0_{\gamma\delta}(i\omega_l,i\omega_m) = - \beta
\left[ \frac{N_c}{N}
  \sum_\mathbf{\tilde{k}} g_{\gamma\delta\downarrow} 
  \left( \mathbf{\tilde{k}}: i\omega_l \right) 
\right]
\left[ \frac{N_c}{N}
  \sum_\mathbf{\tilde{k}} g_{\delta\gamma\uparrow} 
  \left( \mathbf{\tilde{k}}: i\omega_l \right) 
\right] \delta_{l,m}. 
\label{eqn:cf0}
\end{align}
Here, $\hat{C}^0$, $\hat{C}$ and $\hat{\Gamma}$ are 
$N_c N_f \times N_c N_f$ matrices, and 
$N_f$ is the number of the Matsubara frequency. 
On the other hand, 
the bare two-electron Green's function in the lattice system 
is calculated as, 
\begin{align}
C^0_{\gamma\delta}(\mathbf{\tilde{q}}:i\omega_l,i\omega_m) = 
- \frac{\beta N_c}{N} \sum_\mathbf{\tilde{k}} 
g_{\gamma\delta\downarrow} \left( \mathbf{\tilde{k}}+\mathbf{\tilde{q}}: i\omega_l \right)
g_{\gamma\delta\uparrow}   \left( \mathbf{\tilde{k}}                   : i\omega_l \right)
\delta_{l,m}. 
\label{eqn:cf0q}
\end{align}
By using Eqs. (\ref{eqn:vf}) and (\ref{eqn:cf0q}), 
we can compute the lattice two-electron Green's function, 
\begin{align}
\hat{C}\left( \mathbf{\tilde{q}} \right) = 
\left[ {\hat{C^0}\left( \mathbf{\tilde{q}} \right)}^{-1} - \hat{\Gamma} \right]^{-1}. 
\label{eqn:cfq}
\end{align}
Taking account of the phase factor, 
we finally obtain the $\mathbf{q}$-dependent susceptibility, 
\begin{eqnarray}
\chi_{\gamma \delta}(\mathbf{\tilde{q}}) = \frac{1}{\beta^2}
\sum_{l,m} C_{\gamma\delta} \left( \mathbf{\tilde{q}}:i\omega_l,i\omega_m \right)
e^{-i\mathbf{\tilde{q}} \cdot \left( \mathbf{r}_\gamma - \mathbf{r}_\delta \right)}. 
\label{eqn:chi_kl}
\end{eqnarray}

\section{Kagom{\'e} lattice system\label{sec:kagome}}
\begin{figure}[bt]
\begin{minipage}[t]{\minitwocolumn}
\begin{center}
\includegraphics[clip,width=\figsmall]{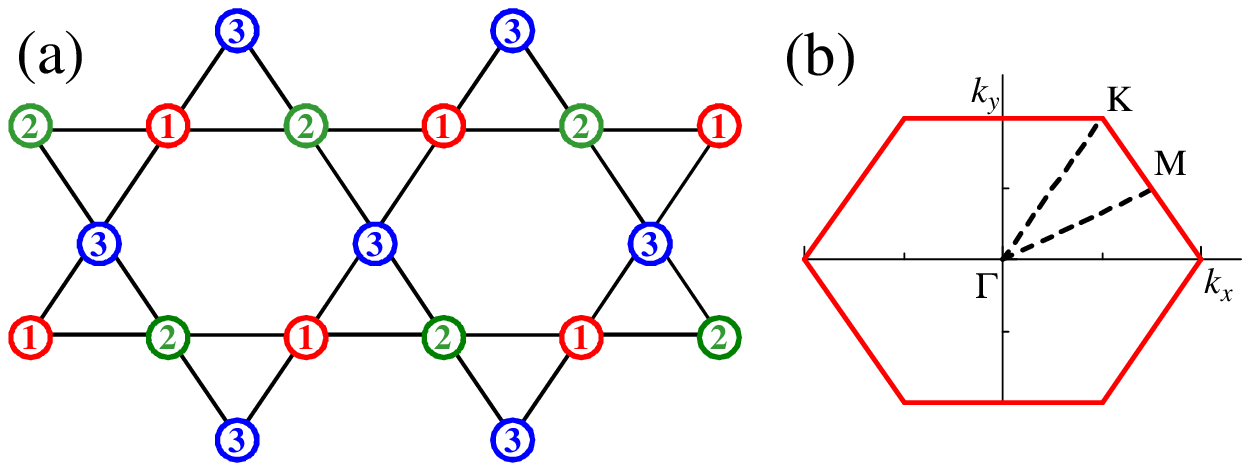}
\end{center}
\caption{\label{fig:kagome}(Color online) 
(a) Sketch of the kagom\'e lattice and  
(b) the first Brillouin zone. }
\end{minipage}
\begin{minipage}[t]{\minispace}
\ 
\end{minipage}
\begin{minipage}[t]{\minitwocolumn}
\begin{center}
\includegraphics[clip,width=\figsmall]{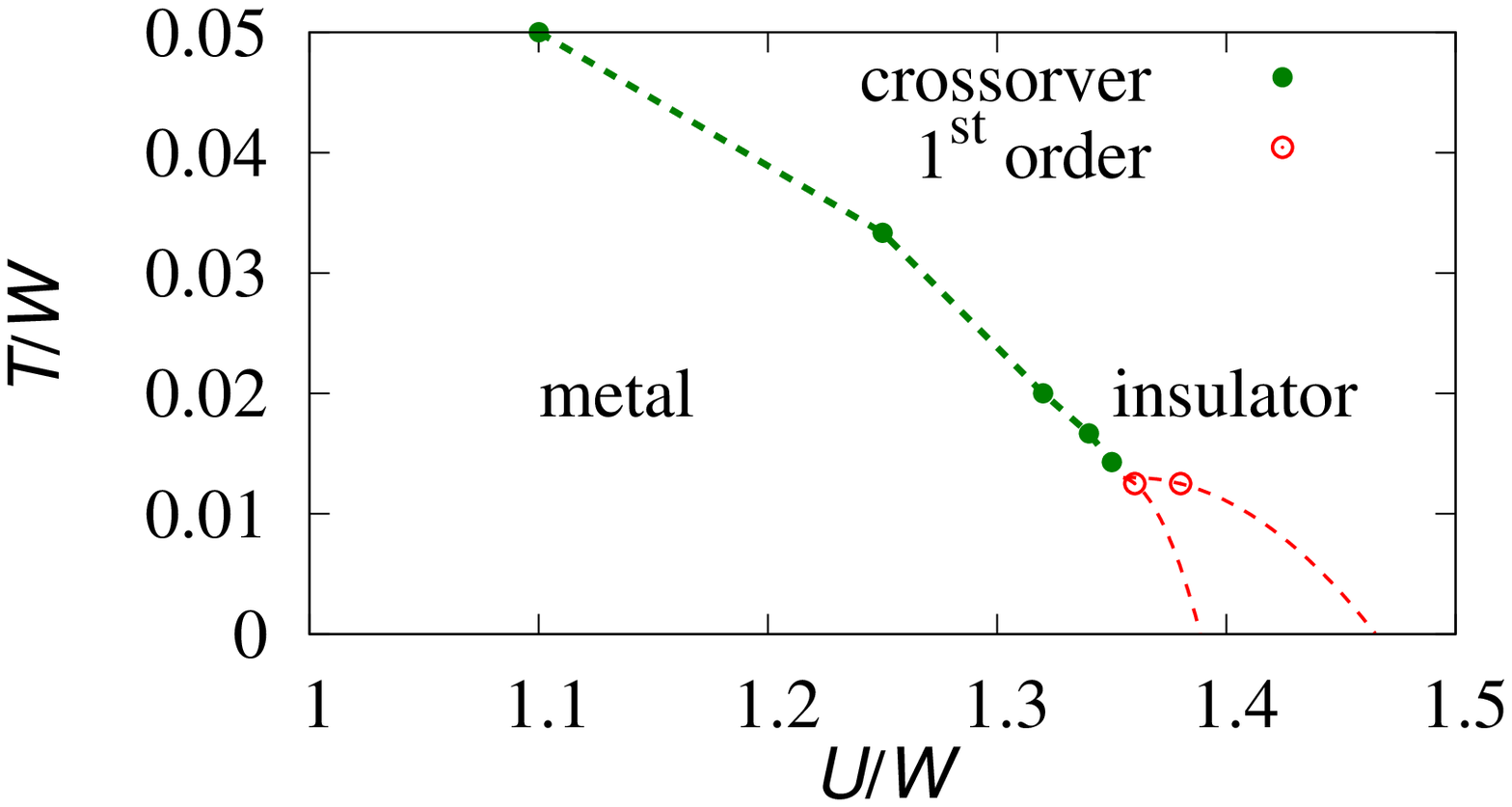}
\end{center}
\caption{\label{fig:phase_kagome}(Color online) 
Phase diagram of the kagom{\'e} lattice Hubbard model. 
The dashed lines are guide to eye. 
}
\end{minipage}
\end{figure}

In this section, we investigate the Mott transition in 
the kagom{\'e} lattice Hubbard model by means of CDMFT
and determine the phase diagram for the Mott transition. 
The result is shown in Fig. \ref{fig:phase_kagome}.
 
Let us consider the Hubbard model (\ref{eqn:hm}) with nearest-neighbor 
hopping on the kagom{\'e} lattice (see Fig. \ref{fig:kagome} (a)), 
\begin{align}
t_{ij}= \left \{
  \begin{array}{lll}
    - t & (t>0) & (\mathrm{site}\ i\ \mathrm{an}\ j:\ \mathrm{nearest\ neighbors}) \\
    0   &       & (\mathrm{otherwise})
  \end{array}
  \right . .
\end{align}
The band width is $W=6t$ and we will use it as an energy unit. 
Unit cell of the kagom\'e lattice has three sites and 
they are labeled by 
$1$, $2$, and $3$, as shown in Fig. \ref{fig:kagome}(a). 
We choose this unit cell as a cluster for the CDMFT approach and 
map the system to an effective cluster model. 
Self-consistent solution of self-energy matrix is obtained 
by means of iterative procedure explained in the previous section, 
and twenty-times iterations are sufficient to achieve satisfactory convergence. 
In each iteration, the local single- and two-electron Green's function 
for the effective model are calculated by QMC, 
where we typically use $10^6$ QMC sweeps and 
Trotter time slices $L = 2W\beta$ 
to reach sufficient computational accuracy. 

\subsection{Mott transition}

\begin{figure}[bt]
\begin{minipage}[t]{\minitwocolumn}
\begin{center}
\includegraphics[clip,width=\figsmall]{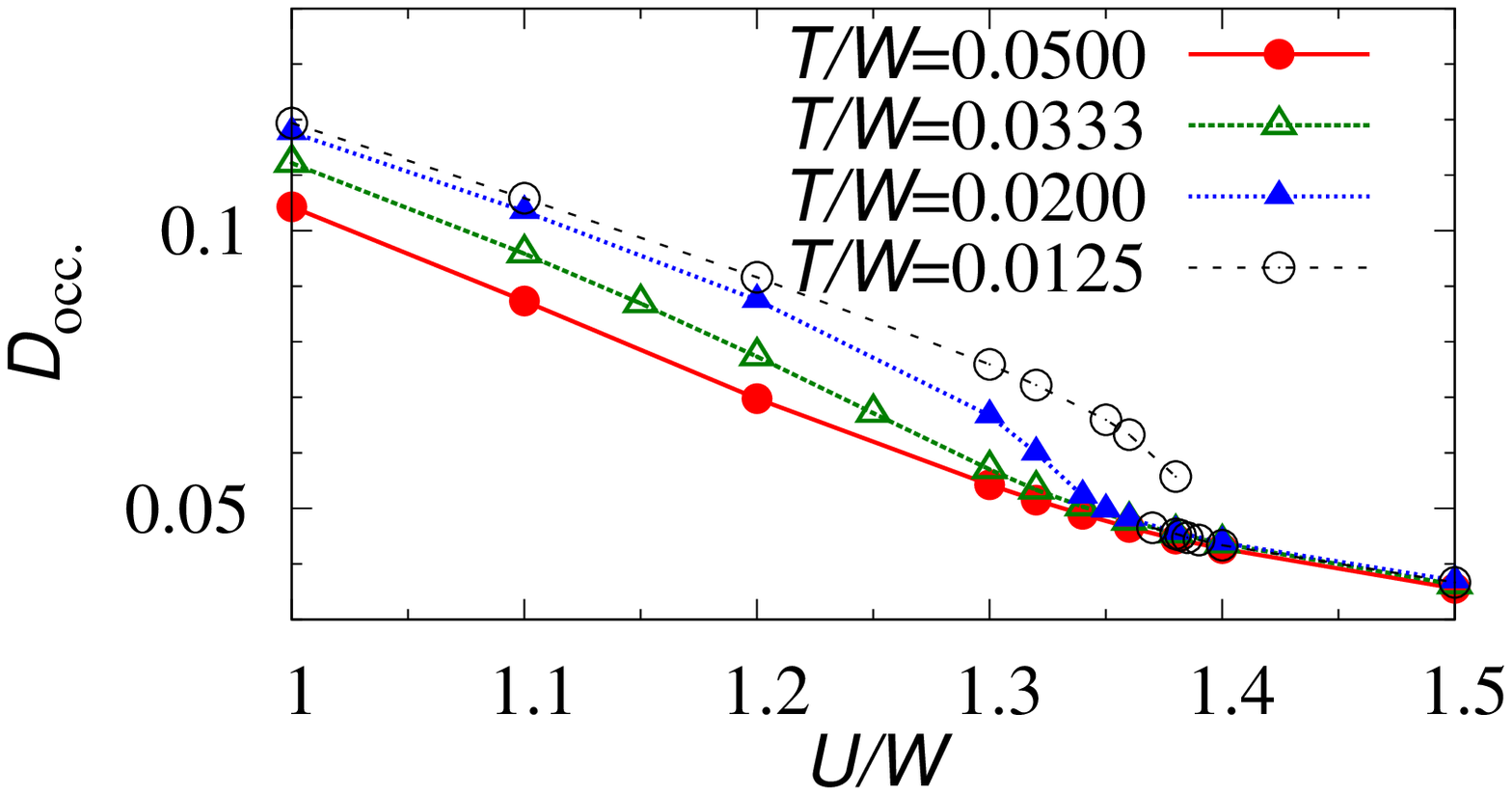}
\end{center}
\caption{\label{fig:double_kl}(Color online) 
Double occupancy as a function of interaction strength $U/W$. 
At $T/W=0.0125$, we can see the discontinuity with hysteresis, 
indicating the first-order Mott transition.
}
\end{minipage}
\begin{minipage}[t]{\minispace}
\ 
\end{minipage}
\begin{minipage}[t]{\minitwocolumn}
\begin{center}
\includegraphics[clip,width=\figsmall]{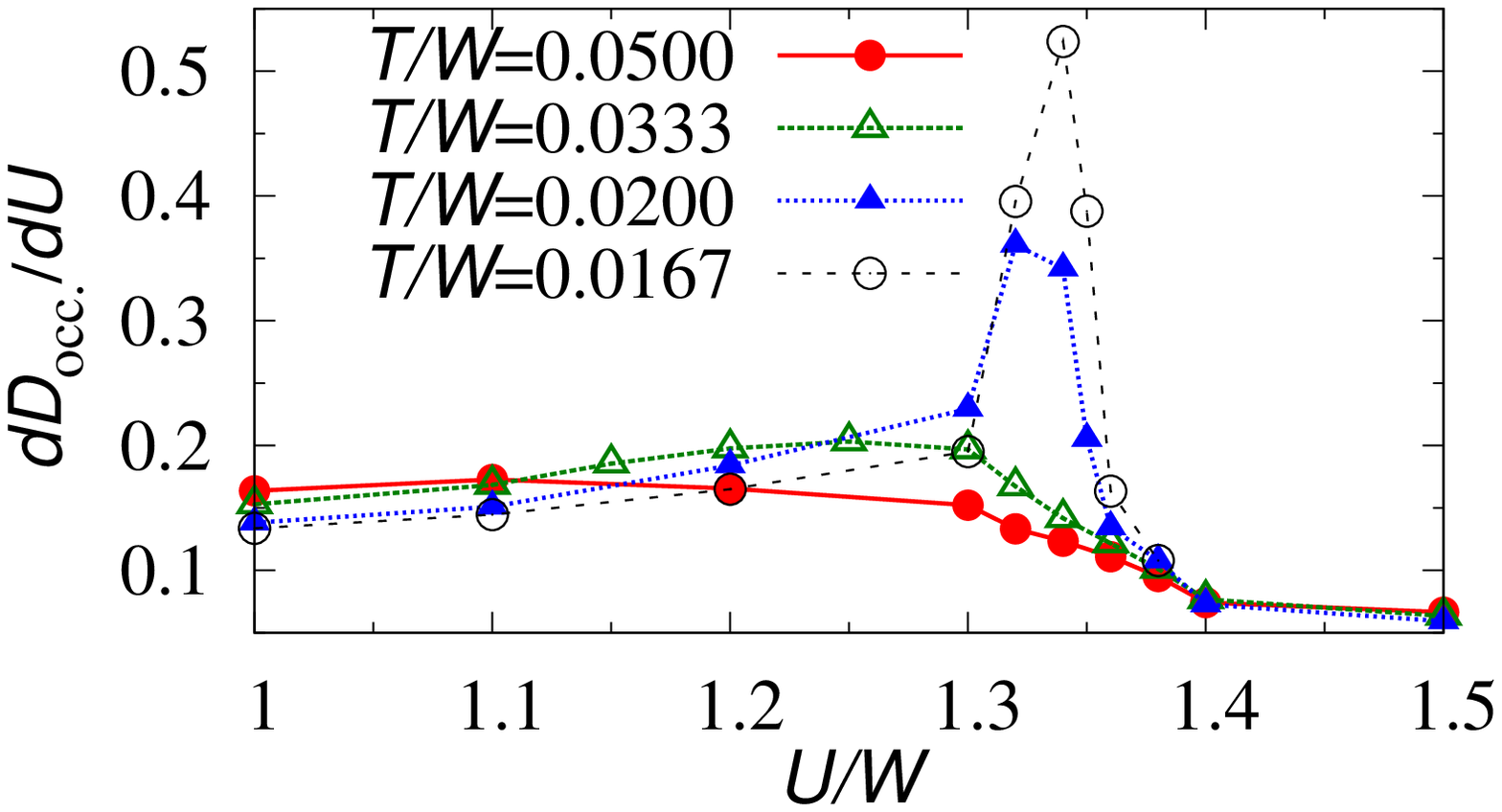}
\end{center}
\caption{\label{fig:derivative}(Color online) 
Derivatives of double occupancy 
$d D_\mathrm{occ.}/d U$. 
We define the crossover point $U^*$ in 
the phase diagram Fig. \ref{fig:phase_kagome} 
by $U$ that gives a maximum of 
$d D_\mathrm{occ. }/d U$. 
}
\end{minipage}
\end{figure}

We now investigate the Mott transition at half filling. 
Figure \ref{fig:double_kl} shows $U$-dependence of the double occupancy 
$D_\mathrm{occ.} = \left \langle n_{i\uparrow} n_{i\downarrow} \right \rangle$ 
for several choices of temperature. At high temperatures, 
$D_\mathrm{occ.}$ smoothly decreases as $U$ increases, which indicates 
that local spin moments are developed. As the temperature
is lowered, a singular behavior emerges around characteristic values of $U$.
When $0.014 \le T/W \le 0.05$, $D_\mathrm{occ. }$ shows a crossover around
$U/W \sim 1.0$-$1.4$. 
For reference, in Fig. \ref{fig:derivative}, we show $d D_\mathrm{occ. }/d U$,
which is computed by numerical differentiation
of $D_\mathrm{occ.}$. 
The maximum of $d D_\mathrm{occ. }/d U$ can be identified as 
the metal-insulator crossover. 
Therefore, we define the crossover point $U^*$ by $U$ that gives
a maximum of $d D_\mathrm{occ. }/d U$. 
At lower temperature $T/W = 0.0125$, the crossover evolves to 
a discontinuity accompanied by hysteresis, which characterizes a 
first-order phase transition at $U_c/W \sim 1.37$. 
We thus end up with the phase diagram shown in Fig. \ref{fig:phase_kagome}, 
where the critical end point is located at $U/W \sim 1.36$ and
 $T/W \sim 0.014$. 
We note that $U_c$ is much larger than 
the crossover strength of $U$ found for the unfrustrated square 
lattice model \cite{moukouri01}. 
As is the case for the triangular lattice \cite{parcollet04}, 
the double occupancy $D_\mathrm{occ.}$ increases in the metallic 
phase ($U<U_c$) as $T$ decreases, 
while it is almost independent of $T$ in the insulating 
phase ($U>U_c$). The increase of $D_\mathrm{occ.}$ at low temperatures
implies that local moments are suppressed due to the
itinerancy of electrons, which in turn leads to 
the development of coherent quasiparticle dynamics. 
It should be noticed that 
in the metallic phase near the critical point, 
$D_\mathrm{occ.}$ starts to increase at very low temperatures. 
This means that the coherence temperature $T_0$ characterizing
quasiparticle formation is very low. 
This naturally gives rise to strong frustration and
 brings about unusual metallic properties near 
the Mott transition, as we will see momentarily below.

\begin{figure}[bt]
\begin{minipage}[t]{\minitwocolumn}
\begin{center}
\includegraphics[clip,width=\figsmall]{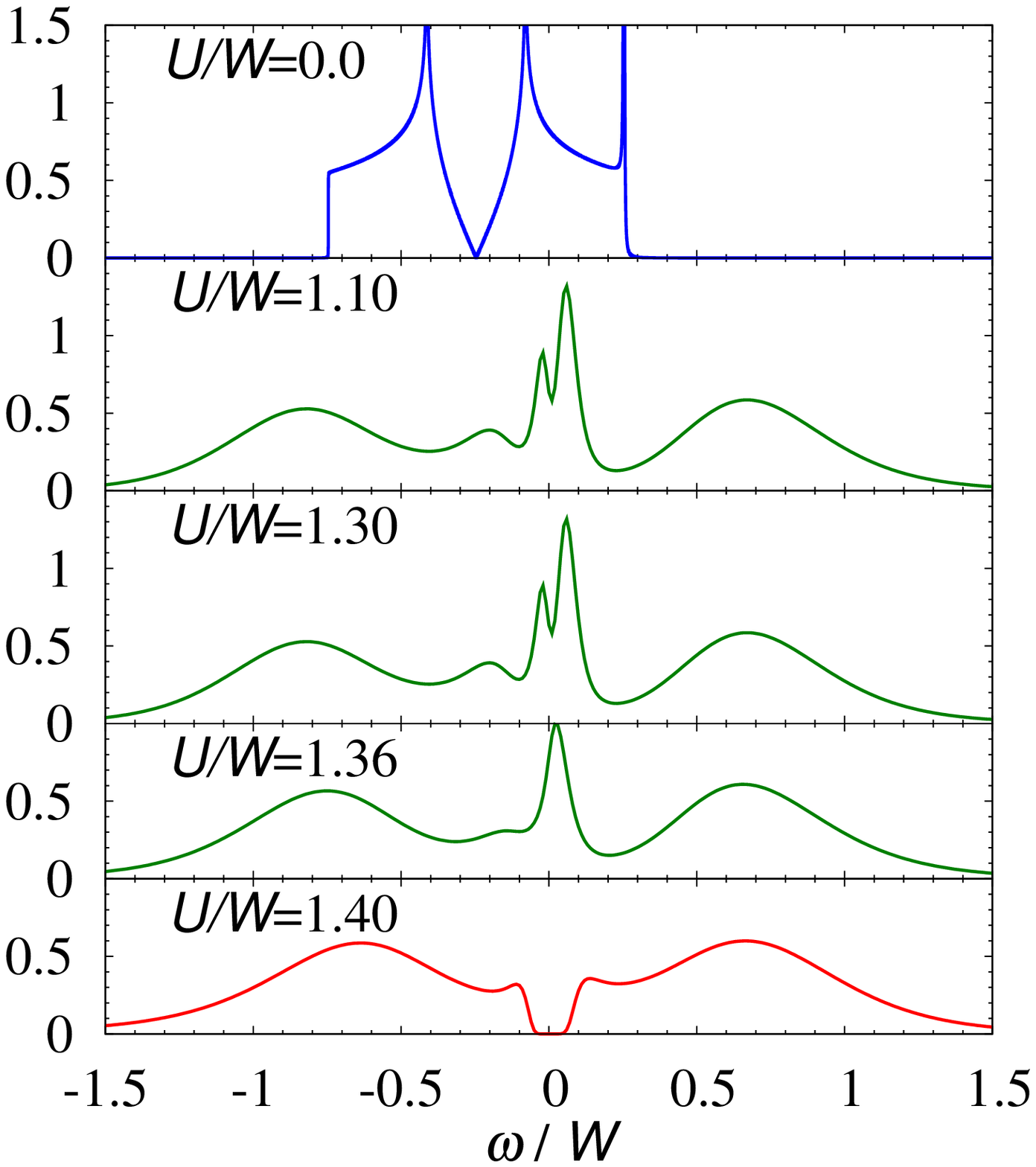}
\end{center}
\caption{\label{fig:dos_kl}(Color online) 
Density of states at $T/W=0.0125$ 
for several values of $U/W$. 
}
\end{minipage}
\begin{minipage}[t]{\minispace}
\ 
\end{minipage}
\begin{minipage}[t]{\minitwocolumn}
\begin{center}
\includegraphics[clip,width=\figsmall]{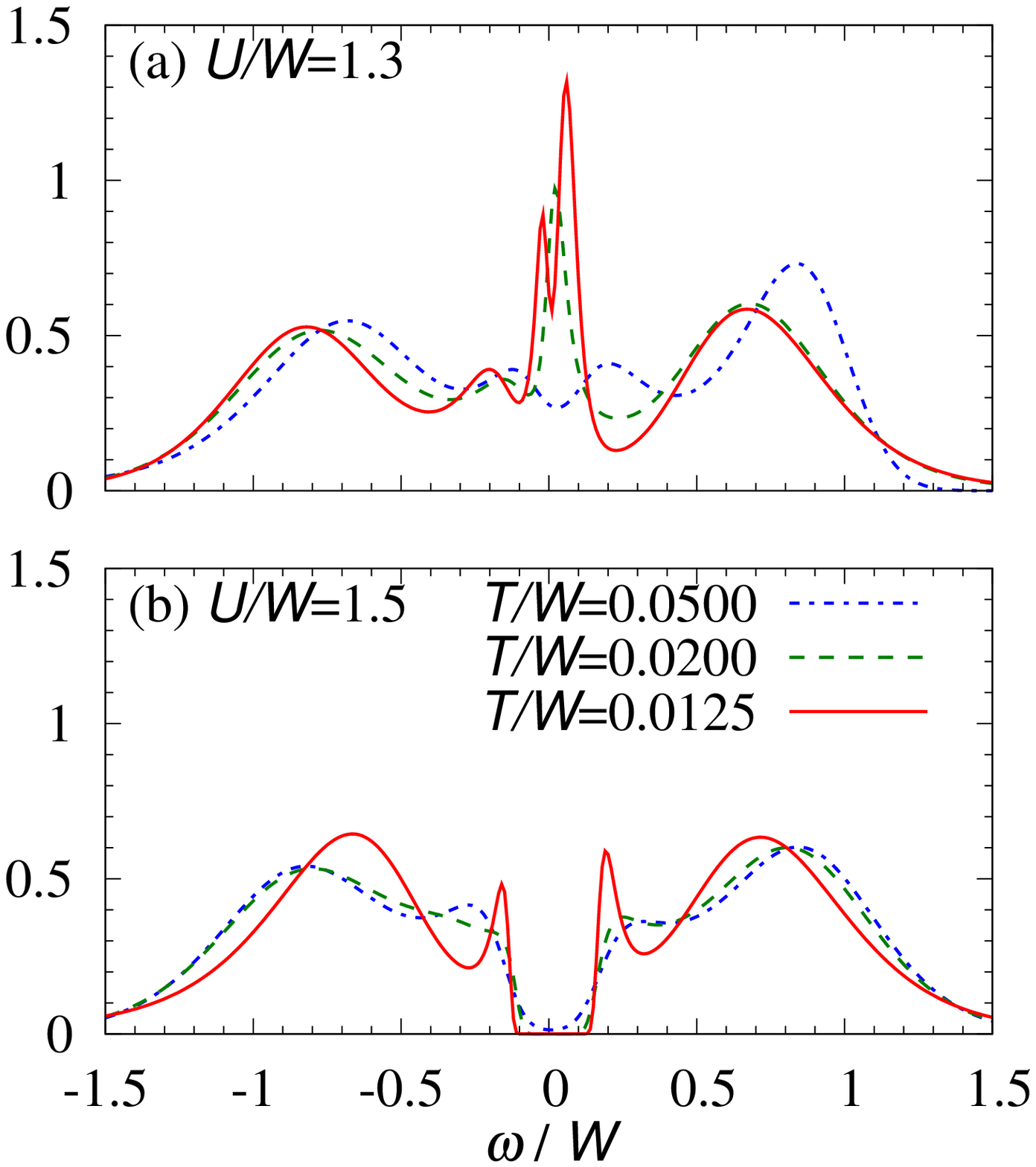}
\end{center}
\caption{\label{fig:dos_kl_t}(Color online) 
Temperature dependence of the density of states 
for $U/W=1.3$ (a) and $U/W=1.5$ (b).
}
\end{minipage}
\end{figure}

In order to clearly examine quasiparticle formation around 
the Mott transition, we compute the density of states (DOS),
$-\frac{1}{\pi}\mathrm{Im}G_{\mathrm{loc}}\left( \omega + i0\right)$. 
This is calculated from the imaginary time single-electron Green's function 
$G_\mathrm{loc}\left( \tau \right)= -\langle \mathrm{T}_\tau 
c_{i\sigma}(\tau) c^\dag_{i\sigma} (0) \rangle$ 
for real frequency $\omega$ by applying MEM. 
In Fig. \ref{fig:dos_kl}, we show DOS at $T/W=0.0125$ for 
several choices of the interaction strength $U/W$. 
At $U=0$, DOS has three distinct energy bands including a 
$\delta$-function peak above the Fermi level.
As $U/W$ increases, DOS forms
heavy quasiparticle peaks around the Fermi level
and finally develops a dip at $U/W \sim 1.40$,
signaling the Mott transition. 
We find two characteristic properties in the metallic phase 
close to the transition point. First, the heavy quasiparticles 
survive up to the transition point ($U/W=1.30$ and $1.36$) and
there is no evidence for pseudo-gap formation, in accordance 
  with the $U$- and $T$-dependence of 
double occupancy in Fig. \ref{fig:double_kl}.
This is related to the suppression of magnetic instabilities 
in our system, in contrast to the square lattice case,
where quasiparticle dynamics are strongly incoherent and 
a pseudo gap opens. The second point is a large renormalization 
of quasiparticle weight near the transition point.
We can see three renormalized peaks near the Fermi level: 
not only the peak near the Fermi surface 
 but also the two other bands away from the Fermi surface are renormalized to
participate in quasiparticle formation. 

Such evolution of quasiparticles can be also clearly seen
in the $T$-dependence of DOS shown in Fig. \ref{fig:dos_kl_t}. 
In the insulating phase $(U/W=1.5)$, 
there is a dip structure near the Fermi level 
already at quite high temperatures and it  
becomes more prominent with lowering $T$, and eventually
a gap opens at low temperatures. 
On the other hand, in the metallic phase close to 
the Mott transition $(U/W=1.3)$, 
the quasiparticle peak develops as $T$ decreases 
instead of the pseudo-gap formation. 
The three quasiparticle peaks evolve near the 
Fermi level with lowering $T$, 
although there exists a dip instead of peak at high temperatures. 
Therefore, the three quasiparticle bands are all relevant
for low-energy excitations near the Mott transition, in contrast to
the weak coupling regime where only the single band 
around the Fermi surface is relevant. 

\subsection{Anomalous spin correlations in the metallic phase}

\begin{figure}[bt]
\begin{minipage}[t]{\minitwocolumn}
\begin{center}
\includegraphics[clip,width=\figsmall]{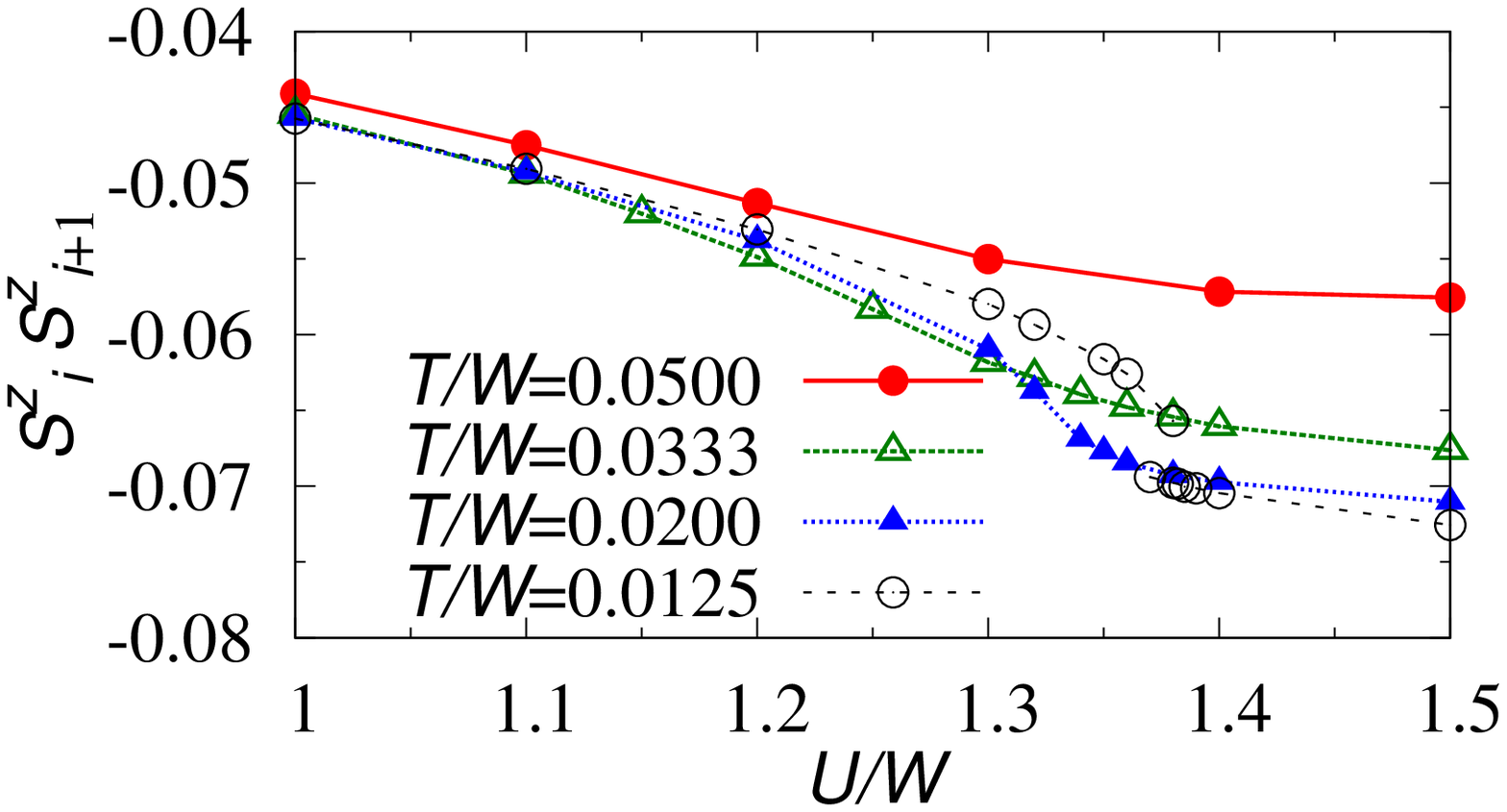}
\end{center}
\caption{\label{fig:spin_u}(Color online) 
Nearest-neighbor spin correlation function 
$\langle S^z_i S^z_{i+1} \rangle$ as a function of $U/W$. 
}
\end{minipage}
\begin{minipage}[t]{\minispace}
\ 
\end{minipage}
\begin{minipage}[t]{\minitwocolumn}
\begin{center}
\includegraphics[clip,width=\figsmall]{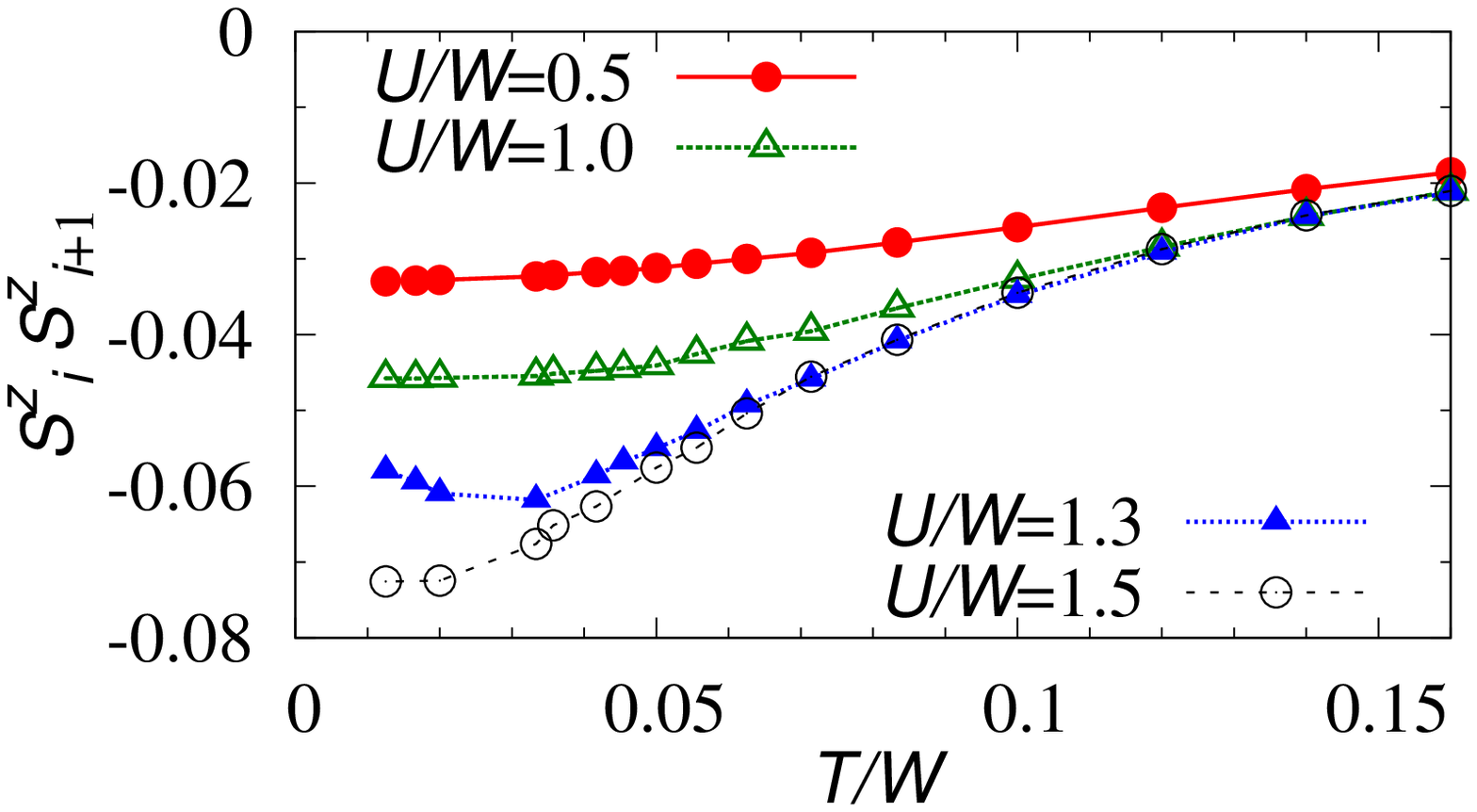}
\end{center}
\caption{\label{fig:spin_t}(Color online) 
Temperature dependence of  
$\langle S^z_i S^z_{i+1} \rangle$ 
for several values of $U/W$.}
\end{minipage}
\end{figure}

It should be noted that quasiparticles exhibit anomalous 
spin correlations due to strong 
frustration around the transition point. 
We show the nearest-neighbor spin correlation function 
$\langle S^z_i S^z_{i+1} \rangle$ at several temperatures
in Fig. \ref{fig:spin_u}. 
Here, 
$S^z_i=( c_{i\uparrow}^\dag c_{i\uparrow} 
- c_{i\downarrow}^\dag c_{i\downarrow} )/2$. 
It is seen that 
$\langle S^z_i S^z_{i+1} \rangle$ is always negative so that 
the spin correlation is antiferromagnetic (AF), 
which gives rise to strong frustration on the kagom{\'e} lattice. 
With increasing  $U/W$, the nearest-neighbor AF spin correlation 
is gradually enhanced. In the insulating phase 
the AF spin correlation becomes stronger as $T$ decreases.
We note that the low-temperature spin correlation 
in the insulating phase is  
weaker than that in isolated triangle, 
$\langle S^z_i S^z_{i+1} \rangle = -1/12$. 
More striking behavior appears in the metallic phase 
near the transition point: the AF spin correlation is once 
enhanced and then suppressed 
with decrease of temperature, as shown in Fig. \ref{fig:spin_t}. 
This anomalous temperature dependence comes from
 the competition between the quasiparticle formation 
and the frustrated spin correlations, which 
is characterized by two energy scales: 
the coherence temperature $T_0$ and 
$T_M$ characterizing the AF spin fluctuations. 
The AF correlation enhanced around $T \sim T_M$
 stabilizes localized moments and 
gives rise to frustration, which is consistent 
 with the monotonic enhancement 
of spin correlations in the insulating phase in Fig. \ref{fig:spin_t}.
On the other hand, 
when the system is in the metallic phase, 
electrons recover coherence in itinerant motion below $T_0$. 
Frustration is thus relaxed by itinerancy of electrons
through the suppression of AF correlations at $T<T_0$. 
Therefore, the nonmonotonic temperature-dependence in 
$\langle S^z_i S^z_{i+1} \rangle$ demonstrates  the
formation of the heavy quasiparticles 
under strong frustration effects. 

\begin{figure}[bt]
\begin{minipage}[t]{\minitwocolumn}
\begin{center}
\includegraphics[clip,width=\figsmall]{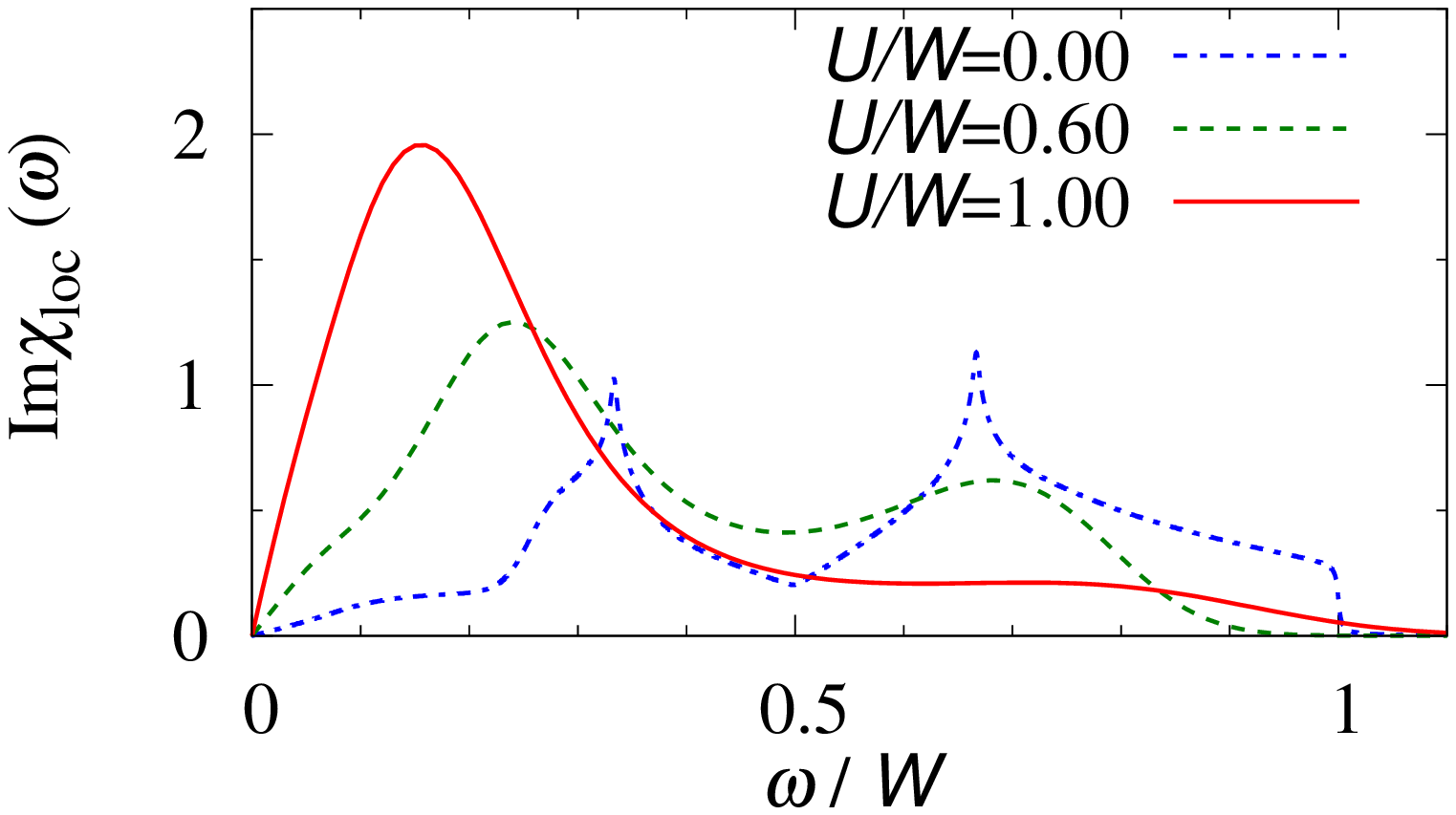}
\end{center}
\end{minipage}
\begin{minipage}[t]{\minispace}
\ 
\end{minipage}
\begin{minipage}[t]{\minitwocolumn}
\begin{center}
\includegraphics[clip,width=\figsmall]{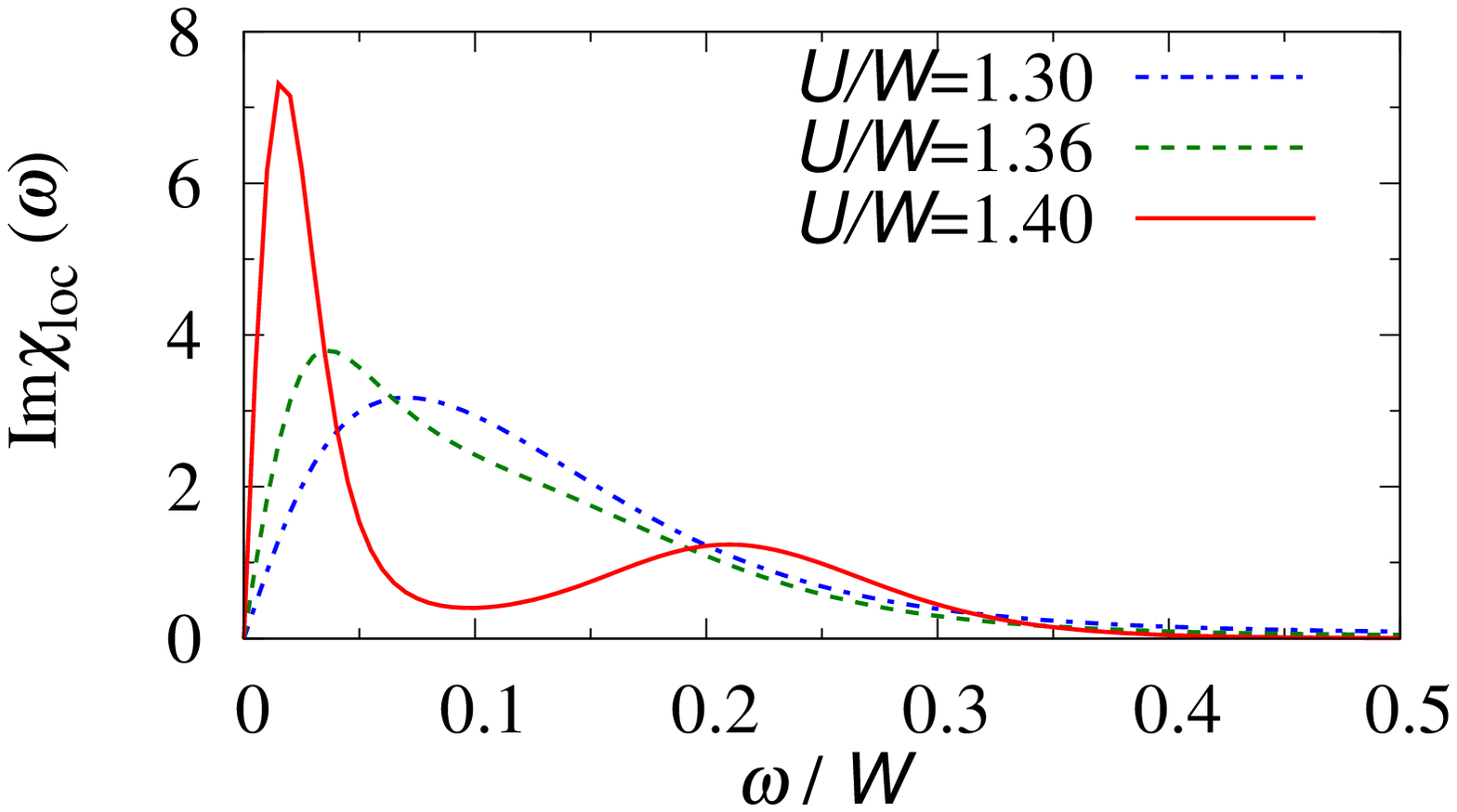}
\end{center}
\end{minipage}
\caption{\label{fig:spin}(Color online) 
Dynamical susceptibility 
$\mathrm{Im} \chi_\mathrm{loc} ( \omega )$ 
for several values of $U/W$ at $T/W=0.0125$. 
}
\end{figure}

We can see the anomalous properties also 
in dynamical spin correlations. 
We compute the dynamical spin susceptibility defined by
\begin{align}
\chi_\mathrm{loc} ( \omega ) = i \int_0^\infty dt e^{i \omega t}
\langle [ S^z_i (t), S^z_i(0)] \rangle. 
\end{align}
Shown in Fig. \ref{fig:spin} is $\mathrm{Im} \chi_\mathrm{loc} ( \omega )$ 
around the Mott transition at $T/W=0.0125$. 
Note that $\mathrm{Im} \chi_\mathrm{loc} ( \omega )$ 
dramatically changes its profile around the Mott transition. 
In the insulating phase ($U/W=1.4$), there are
two distinct peaks  in $\mathrm{Im} \chi_\mathrm{loc} ( \omega )$ at low energies. 
On the other hand, in the metallic phase ($U/W=1.3$, $1.36$), 
two peaks are renormalized into a single peak 
and its peak value is strongly suppressed. 
This is a clear demonstration of drastic change in spin dynamics 
between metallic and insulating phases in frustrated systems. 
The double-peak structure in the insulating phase 
is due to the dominant short-range AF correlations 
at low temperatures. 
The strongly enhanced low-energy peak in $\chi_\mathrm{loc} ( \omega )$ 
corresponds to excitations among the almost degenerate states 
for which a singlet spin pair is formed inside the unit cell, 
while the higher-energy hump is caused by the excitations 
from these low-energy states to other excited states. 
In the metallic phase, the AF correlations are suppressed 
and then frustration is relaxed via the itinerancy of electrons, 
which leads to the renormalized single peak structure 
in $\chi_\mathrm{loc} ( \omega )$. 
We can thus say that the dramatic change in $\chi_\mathrm{loc} ( \omega )$ 
characterizes the competition between itinerancy and frustration of
correlated electrons around the Mott transition. 

\subsection{Enhanced one-dimensional spin correlations}

\begin{figure}[bt]
\begin{center}
\includegraphics[clip,width=\figlarge]{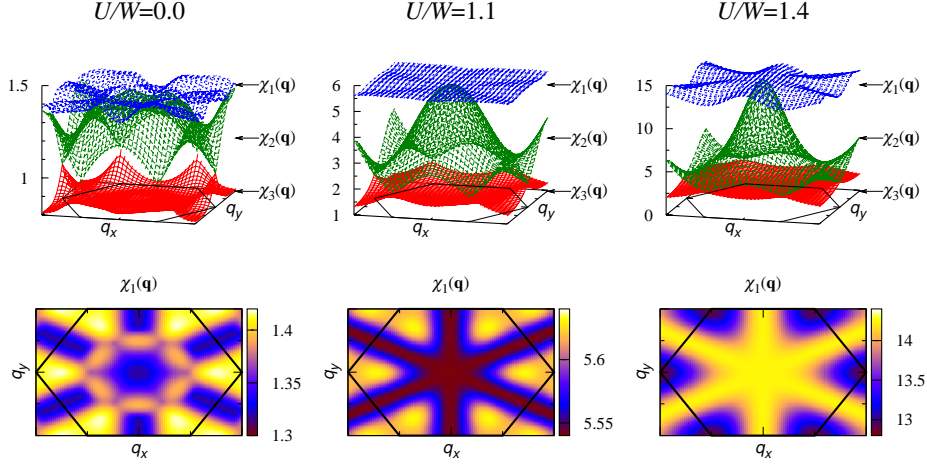}
\end{center}
\caption{\label{fig:chi} 
Wave-vector dependence of the static susceptibility $\chi_{m}(\mathbf{q})$ for several values of $U/W$
at $T/W=0.0333$. The three-dimensional plots of $\chi_{m}(\mathbf{q})$ are shown in the upper panels, 
from top to bottom, $m=1, 2, 3$. The two-dimensional plots in the lower panels show the dominant mode of
the susceptibility $\chi_{1}(\mathbf{q})$ in the upper panels. 
Hexagons in figures denote the first Brillouin zone, as
shown in Fig. \ref{fig:kagome}(b).}
\end{figure}
%
\begin{figure}[bt]
\begin{center}
\includegraphics[clip,width=\figlarge]{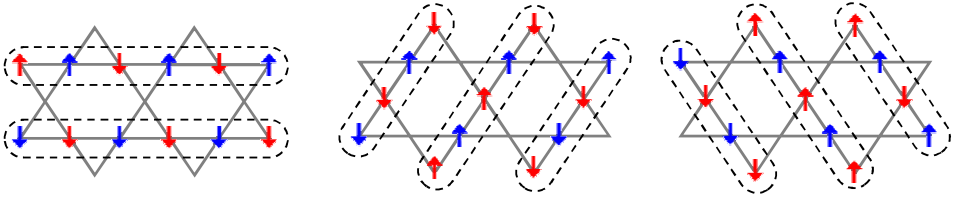}
\end{center}
\caption{\label{fig:config}(Color online) 
Enhanced 1D spin correlations in the insulating phase: 
$U/W=1.4$ at $T/W=0.0333$. }
\end{figure}

We now investigate how these local spin correlations 
affect long range correlations 
by calculating the $\mathbf{q}$-dependence of 
static susceptibility, 
\begin{align}
\chi_{\gamma \delta}(\mathbf{q}) = \int_0^{\beta} d \tau 
\sum_{\mathbf{k},\mathbf{k}'} 
\left \langle 
c_{\mathbf{k}\gamma\uparrow}^\dag               \left ( \tau \right ) 
c_{\mathbf{k}+\mathbf{q}\gamma\downarrow}       \left ( \tau \right )
c_{\mathbf{k}'+\mathbf{q}\delta\downarrow}^\dag \left ( 0 \right )
c_{\mathbf{k}'\delta\uparrow}                   \left ( 0 \right )
\right \rangle ,
\end{align}
where $\gamma,\delta =1,2,3$ denote the superlattice indices. 
We compute $\chi_{\gamma \delta}(\mathbf{q})$, 
following the procedure explained in the previous section, 
where nearest-neighbor as well as local vertex corrections are included. 
Let us introduce $\chi_m(\mathbf{q})$ for three normal 
modes $(\chi_1(\mathbf{q}) > \chi_2(\mathbf{q}) > \chi_3(\mathbf{q}))$ 
defined by eigenvalues of the $3 \times 3$ matrix 
$\chi_{\gamma \delta}(\mathbf{q})$. 
We show  $\chi_{m}(\mathbf{q})$ 
for several choices of $U/W$ at $T/W=0.0333$ in Fig. \ref{fig:chi}. 
The maximum eigenvalue $\chi_{1}(\mathbf{q})$ 
has much weaker $\mathbf{q}$-dependence than that for the other two modes, 
while the second largest mode $\chi_{2}(\mathbf{q})$ has a strong 
$\mathbf{q}$-dependence with a maximum at $\mathbf{q}=(0,0)$. 
We note that these results are in accordance with those obtained 
by FLEX \cite{imai03} and also by QMC approaches \cite{bulut05}. 
However, we find more striking features in the strong coupling regime. 
At $U/W=0$, $\chi_{1}(\mathbf{q})$ has a maximum 
at six points in the Brillouin zone. 
With increase of $U/W$, $\chi_{1}(\mathbf{q})$ is enhanced 
not only at the six points but also on the three lines 
through $\mathrm{\Gamma}$ and $\mathrm{M}$ points. 
Thus the $\mathbf{q}$-dependence of $\chi_{1}(\mathbf{q})$ 
becomes much flatter at $U/W=1.1$ 
than in the $U=0$ case. 
Once the system enters the insulating phase, 
the $\mathbf{q}$-dependence of $\chi_{1}(\mathbf{q})$ dramatically changes 
its character due to the enhancement of short range AF correlations.
At $U/W=1.4$, the susceptibility $\chi_{1}(\mathbf{q})$ 
further grows along the three lines in $\mathbf{q}$ space 
and becomes dominant there 
instead of the six points that give 
the leading magnetic mode in the weak coupling regime. 
Furthermore, the analysis of the eigenvectors of 
$\chi_{1}(\mathbf{q})$ concludes that two spins in the unit cell are 
antiferromagnetically coupled but the other spin is free.
This implies that the enhanced spin fluctuations 
favor a spatial spin configuration in which 
one-dimensional (1D) AF-correlated spin chains 
are independently formed across free spins in three distinct directions. 
We illustrate a schematic picture of 
the three equivalent types of enhanced spin correlations in Fig. \ref{fig:config}. 
This is one of the naturally expected spin correlations 
on the kagom{\'e} lattice, 
since it stabilizes antiferromagnetic configurations in one direction, 
which is more stable than the naively expected spin configuration having 
a singlet pair and a free spin in each cluster. We wish to emphasize that
the 1D correlations found here in the finite-temperature Mott insulating phase 
are different from those 
for the Heisenberg model on the Kagom{\'e} lattice with the nearest-neighbor
 exchange obtained
by both classical and semi-classical approximations 
\cite{harris92,chubukov92}, 
but are similar to the $\mathbf{q}=0$ structure predicted for the 
classical Heisenberg model 
with a further neighbor exchange \cite{harris92}.
The essential difference is that there is almost no correlation 
between the different 
chains in our results for the Hubbard model. 
These 1D correlations have been recently studied  by Udagawa and Motome 
by means of the larger-cluster CDMFT \cite{udagawa08}. They have 
clarified the origin of the 1D correlations in terms of the strong
renormalization effects of electrons at finite temperatures. 

Summarizing this section, we have clarified that the
metallic phase is stabilized up to fairly large $U$
in electrons on the strongly frustrated kagom{\'e} lattice, 
resulting in the three-band heavy quasiparticles. 
We have seen 
that this gives rise to several anomalous properties of spin correlation 
functions in the metallic phase close to the Mott transition point. 
As a characteristic of the kagom{\'e} lattice system, 
novel 1D spin correlations appear 
in the insulating phase at intermediate temperatures. 
With approaching zero temperature, 
the spin liquid state or other nonmagnetic ordered states 
are expected to be realized at low temperatures 
\cite{misguich04,auerbach04}. 

\section{Anisotropic triangular lattice system\label{sec:tri}}

\begin{figure}[bt]
\begin{minipage}[t]{\minitwocolumn}
\begin{center}
\includegraphics[clip,width=0.5\minitwocolumn]{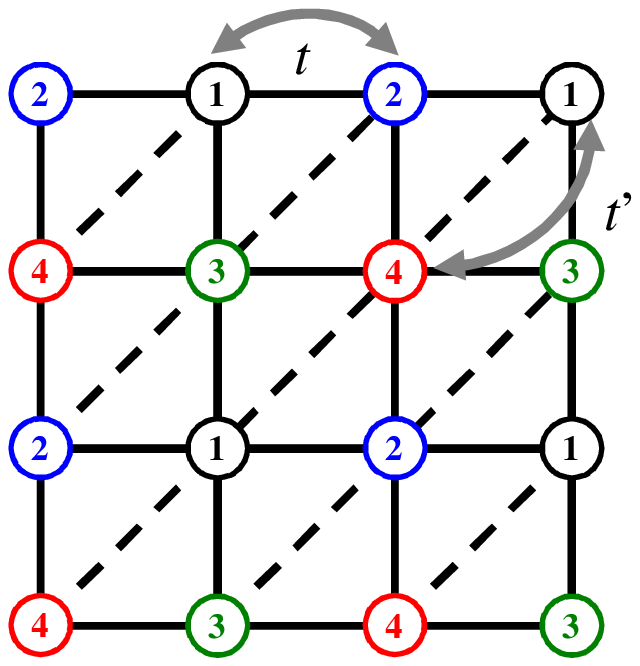}
\end{center}
\caption{\label{fig:lattice}(Color online) 
Sketch of the anisotropic triangular lattice
}
\end{minipage}
\begin{minipage}[t]{\minispace}
\ 
\end{minipage}
\begin{minipage}[t]{\minitwocolumn}
\begin{center}
\includegraphics[clip,width=\figsmall]{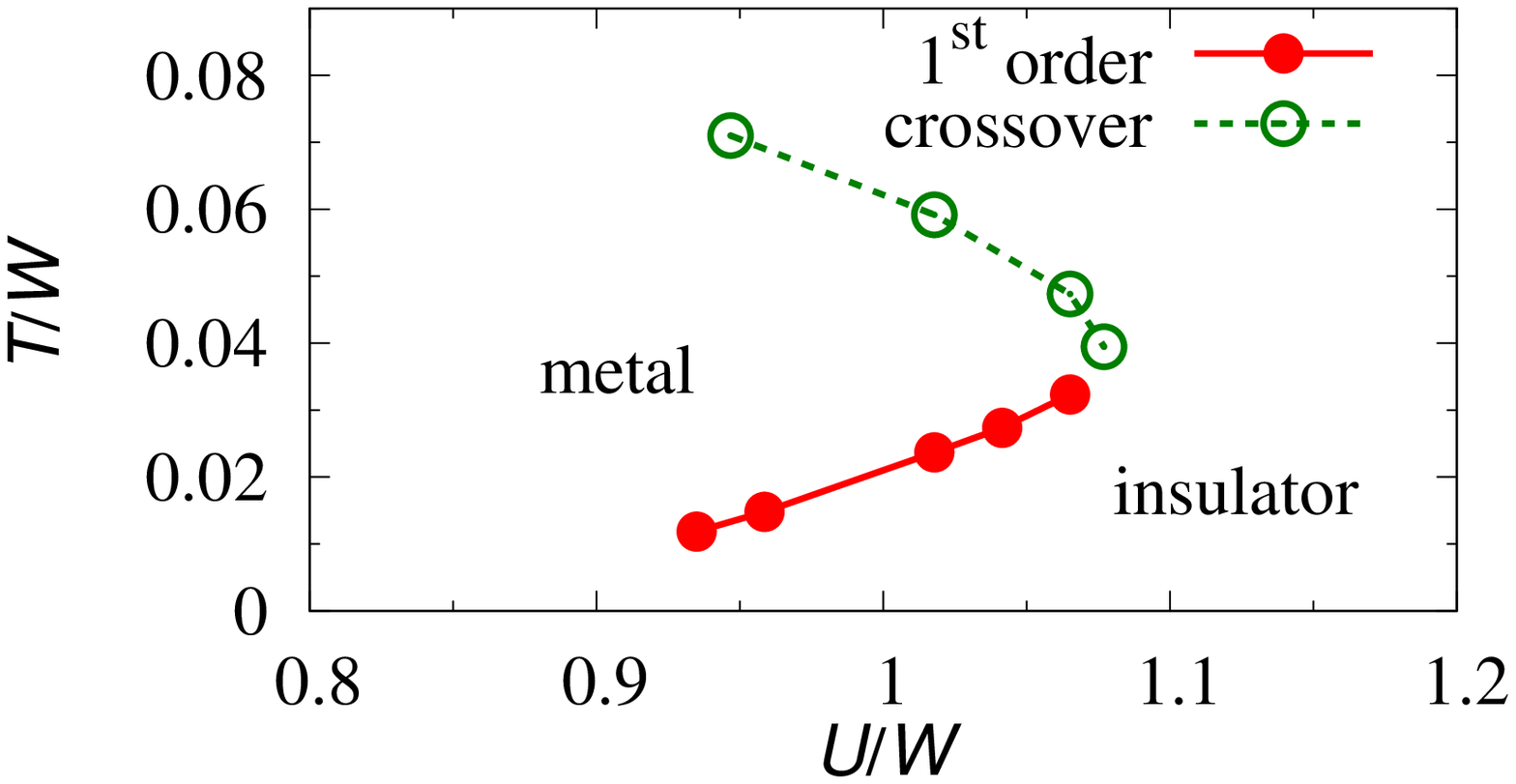}
\end{center}
\caption{\label{fig:phase}(Color online) 
Phase diagram of Hubbard model on 
anisotropic triangular lattice for $t'/t=0.8$.
}
\end{minipage}
\end{figure}

In this section, we study the finite-temperature Mott transition
in the Hubbard model on the anisotropic triangular lattice. 
We determine the $U$-$T$ phase diagram as shown 
in Fig. \ref{fig:phase}, which shows a remarkable property,
{\it i.e.} reentrant behavior in the Mott transition.  

Let us consider the Hubbard model (\ref{eqn:hm}) 
on the anisotropic triangular lattice shown in Fig. \ref{fig:lattice}, 
\begin{align}
t_{ij}= \left \{
  \begin{array}{lll}
    - t  & (t >0) & (\mathrm{site}\ i\ \mathrm{an}\ j:\ \mathrm{nearest\ neighbors}) \\
    - t' & (t'>0) & (\mathrm{site}\ i\ \mathrm{an}\ j:\ \mathrm{diagonal}) \\
    0    &        & (\mathrm{otherwise})
  \end{array}
  \right . .
\end{align}
We use the band width $W$ as an energy unit for given anisotropy $t'/t$, 
and it is $W=4(t+t')+t^2/t'$ for $t'>0.5t$ and $W=8t$ for $t'<0.5t$. 
To analyze this model, we use the four-site cluster CDMFT. 
Considering four sublattices labeled by $1$-$4$, 
as shown in Fig. \ref{fig:lattice}, 
we map the original lattice model to a four-site cluster model 
coupled to the self-consistently determined medium. 
Fifty-times iterations in a CDMFT procedure 
and typically $10^6$ QMC sweeps with Trotter time slices 
$L = 12t\beta$ are performed to obtain numerical convergence. 

\subsection{Mott transition}

\begin{figure}[bt]
\begin{center}
\includegraphics[clip,width=1.54\figsmall]{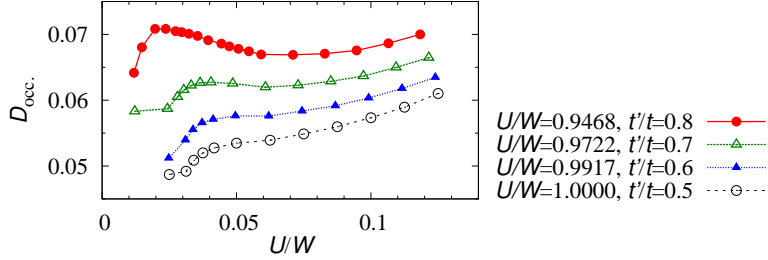}
\end{center}
\caption{\label{fig:double}(Color online) 
Nonmonotonic temperature dependence of double occupancy 
in the typical strong-$U$ regime $U \sim W$, 
where ground states are insulating and in the
vicinity of the Mott transition. 
}
\end{figure}

We investigate the Mott transition for the model
(\ref{eqn:hm}) at half filling. To this end, let us first compute the
$T$-dependence of the double occupancy $D_\mathrm{occ.} = \left
\langle n_{i\uparrow} n_{i\downarrow} \right \rangle$ for 
typical interaction strength $U/W$, 
where the ground state is insulating 
and also in the vicinity of Mott transition point. 
We find a remarkable property in our frustrated system, 
{\it i.e.} nonmonotonic $T$-dependence 
of $D_\mathrm{occ.}$. It is seen in
 Fig.\ \ref{fig:double} that $D_\mathrm{occ.}$
decreases at high temperatures, and then exhibits an upturn 
in the intermediate temperature region 
having a local minimum at $T/W \sim 0.06$, as $T$
decreases. At much lower temperatures, $D_\mathrm{occ.}$ 
starts to decrease again and thus shows a hump structure. 
The nonmonotonic $T$-dependence of $D_\mathrm{occ.}$ 
implies that our
system once changes from insulating to metallic and then
reenters the insulating phase as $T$ decreases. This
 is quite different from that known for the infinite dimensional
Hubbard model, where $D_\mathrm{occ.}(T)$ has a
single minimum at the Fermi-liquid coherence temperature $T_0$. 
In the latter model, the system is insulating at $T>T_0$ and 
Fermi liquid like at $T<T_0$ \cite{georges96}. 
The nonmonotonic behavior found here is 
also different from that for the unfrustrated square lattice Hubbard model. 
In the square-lattice Hubbard model, 
the Fermi-liquid coherence is 
disturbed by the antiferromagnetic (AF) interaction due to the
perfect nesting of Fermi surface, 
which results in monotonic decrease of $D_\mathrm{occ.}$
\cite{moukouri01,maier05}. 
It is seen that the hump structure in $D_\mathrm{occ.}$ becomes more prominent 
and shifts to lower temperatures as $t'$ increases, 
although it is less visible for $t'/t=0.5$. 

\begin{figure}[bt]
\begin{center}
\includegraphics[clip,width=1.7\figsmall]{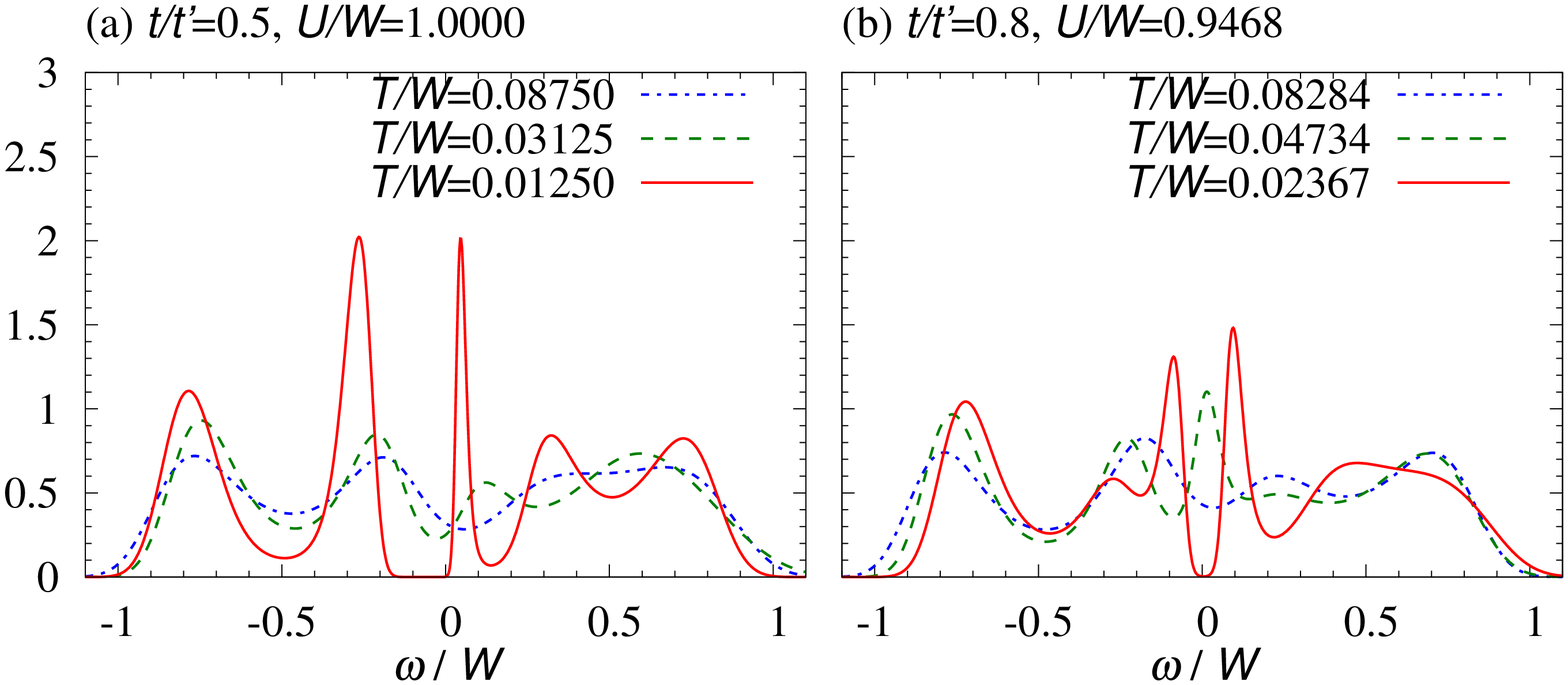}
\end{center}
\caption{\label{fig:dos}(Color online) 
Temperature dependence of DOS for $t'/t=0.5$ (a) 
and $t'/t=0.8$ (b). 
For $t'/t=0.8$, the metallic quasiparticle peak appears 
near the Fermi level at the intermediate temperature
$T/W=0.04734$. 
}
\end{figure}

We can see such characteristic 
behavior also in DOS as shown in Fig. \ref{fig:dos}. 
This is wave-vector-integrated electron spectral function. 
For $t'/t=0.5$, DOS shows a dip structure near the
 Fermi level at high temperatures. 
As $T$ decreases, the dip becomes prominent, 
and finally a gap opens clearly in the low-$T$ insulating phase. 
On the other hand, for $t'/t=0.8$ with stronger frustration, 
 the nonmonotonic behavior appears in DOS. 
The quasiparticle peak develops near the Fermi level 
with lowering temperature, 
although the dip structure appears at high temperatures. 
As $T$ further decreases, the quasiparticle peak disappears and 
an insulating gap opens again. 
These properties are consistent with the results of $D_\mathrm{occ.}$. 
Therefore, it is concluded that the nonmonotonic behavior is a characteristic 
feature caused by geometrical frustration. 

\begin{figure}[bt]
\begin{center}
\includegraphics[clip,width=\figsmall]{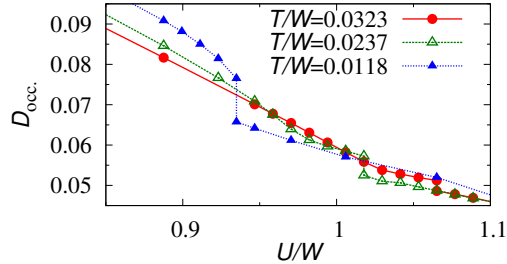}
\end{center}
\caption{\label{fig:transition}(Color online) 
Double occupancy as a function of interaction strength
$U/t$ for $t'/t=0.8$ at several temperatures. 
We show only the transition from insulator to metal with weakening $U$ 
(defined as $U_{c1}$), although we find hysteresis. 
}
\end{figure}

In order to clarify whether the change between  metal and 
insulator is a real phase transition or crossover, we examine the
double occupancy for typical anisotropy $t'/t=0.8$ with varying $U$. 
Let us start from the noninteracting system to reach the
large-$U$ regime, typically $U/W \sim 1.2$, and then calculate
$D_\mathrm{occ.}$ with gradually decreasing $U$. 
It is seen in Fig. \ref{fig:transition} that
the double occupancy jumps at critical 
interaction strength $U_c$ as $U$ decreases, signaling a
first-order Mott transition. 
The size of the jump shrinks as $T$ increases, and eventually
vanishes above $T/W \sim 0.035$. It is expected that
the critical end point is located at  
$T/W \sim 0.035$ and $U/W \sim 1.077$. 
At high temperatures, the system exhibits 
a crossover between metal and insulator, where we define the
boundary by the temperature at which the double occupancy takes the
first local minimum at high temperatures, as seen in Fig. \ref{fig:double}. 
Note that the metal-insulator boundary is in accordance with that determined
by the local minimum of the density of states at the Fermi energy. 
We thus end up with the phase diagram shown in Fig. \ref{fig:phase}. 
We wish to emphasize here that $U_c(T)$ in our system 
has a slope with the opposite sign to the behavior in the infinite
dimensional model at low temperatures, whereas 
the high-$T$ crossover exhibits similar behavior to 
the infinite-dimensional case. 

\subsection{Reentrant behavior}

It is known that the single-site DMFT gives
entropy of about $\ln 2$ per site in the paramagnetic insulating 
phase corresponding to localized free spins 
and smaller entropy is realized 
in the metallic phase with the Fermi-liquid coherence. 
Therefore, near the Mott
transition temperature, the system is insulating at high
temperatures to gain large entropy and the metallic phase exists 
always in the lower temperature region \cite{georges96}. 
It should be noted, however, that the spatial fluctuations, 
which are not taken into account in the single-site DMFT, 
are important at low temperatures. 
For instance, according to the dynamical cluster study of
 the square-lattice Hubbard model, the
Fermi-liquid metallic phase does not appear 
because of strong AF correlations \cite{moukouri01,maier05}. 
In contrast,  the magnetic correlations in our system are 
hard to develop until low temperature 
$T/W\sim 0.05$ due to strong geometrical frustration. Hence,
when the temperature is lowered, 
the entropy is released not by spin correlations but
by the itinerancy of electrons at $T/W>0.05$, which 
gives rise to  the crossover behavior 
from insulator to metal as shown in Fig. \ref{fig:phase}.
The emergence of such Fermi-liquid states is one 
of the characteristics in the vicinity of  the
Mott transition with geometrical frustration \cite{ohashi06}. 
Note that the magnetic correlations are enhanced 
at much lower temperatures, 
finally triggering a first-order phase transition from 
the Fermi liquid to an insulator with smaller entropy. 
Hence with  decreasing $T$,  $U_c(T)$ decreases at 
low temperatures ($T/W<0.05$) 
in contrast to the behavior at higher temperatures ($T/W>0.05$). 
We can thus say that the reentrant Mott transition found here for
the anisotropic  triangular lattice 
is caused by the competition induced by geometrical frustration 
between the Fermi-liquid formation and the magnetic correlations.

\begin{figure}[bt]
\begin{center}
\includegraphics[clip,width=1.3\figsmall]{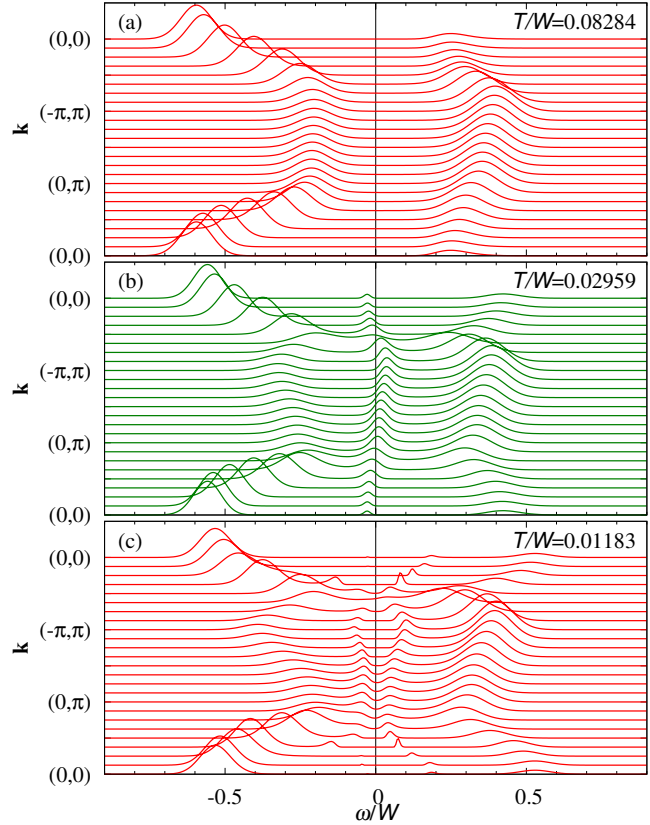}
\end{center}
\caption{\label{fig:arpes}(Color online) 
Momentum resolved single-electron spectrum $A_\mathbf{k}(\omega)$
for $U/W=0.9468$, $t'/t=0.8$ at several temperatures.
}
\end{figure}

Let us now discuss the momentum resolved single-electron spectrum
$A_\mathbf{k}(\omega)$, in which 
we can clearly see the development of the quasiparticles and 
magnetic correlations discussed above. 
This is calculated by using Eq. (\ref{eqn:green}) and MEM. 
We show $A_\mathbf{k}\left( \omega \right)$ for $U/W=0.9468$, 
$t'/t=0.8$ at typical temperatures
in Fig. \ref{fig:arpes}. It is seen that
$A_\mathbf{k}\left( \omega \right)$ 
shows an insulating behavior at high temperatures, where it has
a large Hubbard gap of order of $U/W$ and no quasiparticle peak.
With decreasing temperature, a quasiparticle peak starts to develop 
inside the gap, and results in a quasiparticle band with weak dispersion. 
This clearly suggests the appearance of the frustration-induced 
metallic phase. 
The emergence of the metallic phase due to geometrical frustration is 
in accordance with the previous studies of the Hubbard model on the triangular 
lattice \cite{imai02,maier05}
and the kagom{\'e} lattice \cite{ohashi06}.
As temperature further decreases, the quasiparticle peaks split and 
acquire a very small gap, and the system enters another insulating phase again. 
Note that the small gap is caused by the  exchange interaction
among quasiparticles, which is consistent with the results 
of  the CDMFT study  with exact 
diagonalization method at zero temperature\cite{kyung06prl}. 
It is thus confirmed that the present frustrated system 
exhibits the insulator-metal-insulator reentrant behavior
with decreasing temperature. 

\subsection{Magnetic instability}

\begin{figure}[bt]
\begin{center}
\includegraphics[clip,width=\figlarge]{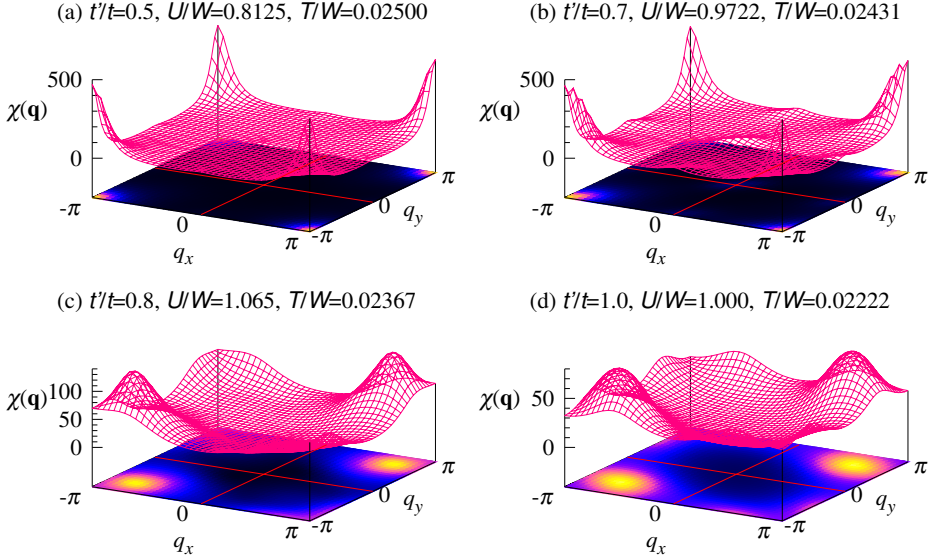}
\end{center}
\caption{\label{fig:sus}
Static spin susceptibility $\chi(\mathbf{q})$ around 
the Mott transition point. 
}
\end{figure}

Here, we wish to briefly discuss magnetic instability. 
The static spin susceptibility $\chi(\mathbf{q})$ 
computed for typical $U/W$ and $T/W$ around the Mott transition
is shown in Fig. \ref{fig:sus}. 
For $t'/t<0.7$ (weak frustration), it is seen that 
$\chi(\mathbf{q})$ develops a peak structure with lowering $T$
 at $\mathbf{q} = (\pi, \pi)$ corresponding 
to the commensurate antiferromagnetism.
The behavior is consistent with recent studies of
the Hubbard model around the Mott 
transition \cite{kashima01,yokoyama06,koretsune07}. 
Quite different tendency appears at $t'/t=1.0$ (triangular lattice), 
where the spin susceptibility 
$\chi(\mathbf{q})$ has a maximum at $\mathbf{q} = (2 \pi/3, 2 \pi/3)$ 
corresponding to the $120^\circ$ structure, and
the development of $\chi(\mathbf{q})$ is much slower than 
the AF correlations for small $t'$. 
For $U/W=1.065$, $t'/t=0.8$ at $T/W=0.02367$, 
where the system is in the insulating phase near
the  Mott transition point, 
 $\chi(\mathbf{q})$ does not diverge, 
but takes a maximum at incommensurate wave vectors
$\mathbf{q} \sim (0.7 \pi, 0.7 \pi)$.
It is remarkable that $\chi(\mathbf{q})$ remains finite just below 
the first order Mott transition temperature and hence
the paramagnetic Mott insulator is not precluded by the 
magnetically ordered phase. This is  
contrasted to the single-site DMFT results: the magnetic order
conceals the Mott transition \cite{georges96,zitzler04}. 
On the other hand, 
the magnetic ordering in our case, 
which is strongly suppressed by frustration effects, 
emerges below the Mott transition temperature,
although a finite-temperature magnetic transition is 
due to a mean-field type approximation. 
In this way, the magnetic correlations play a crucial
role in driving the Mott transition at low temperatures, 
although they do not trigger a real magnetic instability. 

\subsection{Controlling frustration}

It is to be noted that the determined phase diagram (Fig. \ref{fig:phase}) in 
the $U$-$T$ plane for $t'/t=0.8$ 
is qualitatively consistent with the experimental data \cite{kagawa04} 
in $\kappa$-(BEDT-TTF)$\mathrm{_2Cu[N(CN)_2]Cl}$ with $t'/t \sim 0.75$.
Now, a question naturally arises: what happens if we control frustration 
by changing the ratio of $t'/t$.  To this end,
we have systematically studied the Mott transition 
for different choices of $t'/t$. 
For weakly frustrated case $t'/t \le 0.5$, we have found that 
the Fermi-liquid formation is suppressed by the strong AF correlations, and
thus the reentrant Mott transition becomes less prominent. 
In this case, the Mott transition temperature is very low \cite{park08} and 
the AF long range order hides the Mott transition within CDMFT approach. 
For fully frustrated case $t' \sim t$, on the other hand,
the Fermi-liquid states are well stabilized by frustration and 
the metallic region is extended, 
because the magnetic fluctuations of the $120^\circ$ structure are weak. 
We thus expect that the low-temperature Mott transition line 
shifts to lower-temperature  regime 
\cite{parcollet04}.  Note that the above tendency is consistent with 
the experiments on another organic material 
$\kappa$-(BEDT-TTF)$_2\mathrm{Cu_2(CN)_3}$ with $t' \sim t$,\cite{shimizu03}
where any magnetic order was not observed so far experimentally.
We conclude that the reentrant behavior can be  
observed  most clearly in electron systems with moderate frustration. 

Summarizing this section, we have found  novel 
 reentrant behavior in the Mott transition in the anisotropic 
triangular lattice Hubbard model, and have clarified that 
the reentrant behavior in the Mott transition is 
caused by the competition 
between the Fermi-liquid formation and the magnetic correlations
under geometrical frustration.

\section{Summary\label{sec:summary}}

We have studied the Hubbard model on 
the kagom{\'e} and anisotropic triangular lattice
by means of the cellular dynamical mean field theory. 
Through the systematic studies, we have elucidated some 
characteristic properties
common to frustrated electron systems near the Mott transition.
First, we have shown that the metallic phase is stabilized up to 
fairly large Hubbard interactions 
under strong frustration in both models. This naturally leads to
the formation of strongly renormalized heavy fermions near the 
Mott transition point, where some 
anomalous properties are caused by almost localized 
electrons with strong geometrical frustration.
One of such anomalous features emerges in the nonmonotonic temperature 
dependence  of the spin correlation function found for
the  kagom{\'e} lattice  model around the Mott transition. 
This is indeed due to electron correlations strongly 
influenced by frustration.
We have also found that such anomalous spin correlations 
result in  more striking behavior in the Mott transition 
in the case of the anisotropic triangular lattice system.
Namely, the Mott transition shows novel reentrant behavior 
due to the competition between Fermi-liquid formation and magnetic correlations, 
as typically seen for the triangular lattice system for $t'/t=0.8$. 

Although we have used the small-cluster CDMFT in this paper, 
we believe that the anomalous properties found here for the Mott transition 
are robust, which may not qualitatively change even for a 
larger cluster size, since the competition between the Fermi-liquid formation and 
magnetic correlations should occur 
generally in frustrated electron systems,
Therefore, we expect that nonmonotonic temperature dependence of 
the spin correlation function and/or the reentrant behavior in 
the Mott transition  will be observed experimentally in a variety 
of frustrated  electron systems.

\section*{Acknowledgment}
We are deeply indebted to our collaborators in this field, 
K. Inaba, A. Koga, Y. Motome, S. Suga and have benefited from helpful discussions
with T. Koretsune, M. Udagawa, R. Arita, S. Onoda, and M. Imada. 
Discussions during the YITP workshop YKIS2007 on 
"Interaction and Nanostructural Effects in Low-Dimensional Systemsh 
were useful to complete this work. 
This work was partly supported by Grants-in-Aid for Scientific Research 
(No. 17071011, No. 19052003, No. 20029013, No. 19014013 and No. 19840031) 
and also by the Next Generation Supercomputing Project, Nanoscience Program, 
from the Ministry of Education, Sports, Science and Culture of Japan. 
A part of numerical computations was done at the Supercomputer Center 
at the Institute for Solid State Physics, 
University of Tokyo and Yukawa Institute Computer Facility.


\end{document}